\documentclass{article}
\usepackage[margin=1in]{geometry}
\usepackage[english]{babel}
\usepackage{graphicx}
\usepackage{amsmath,amssymb}
\usepackage{appendix}
\usepackage{comment}
\usepackage{physics}
\usepackage{hyperref}
\renewcommand{\l}[1]{^{(#1)}}
\newcommand{\lT}[1]{^{(#1)T}}
\newcommand{\evsq}[3]{\left \langle #1 \right \rangle_{#2;#3}}
\newcommand{\Zsq}[2]{\mathcal Z_{#1;#2}}
\newcommand{\kl}[2]{\textrm{KL}\left(#1|#2\right)}
\newcommand{\qb}{Q^{i}_{\alpha\beta}}
\newcommand{\sigb}{\Sigma^{i}_{\alpha\beta}}

\newcommand{\gives}{\rightarrow}
\newcommand{\E}[1]{\mathbb E\left(#1\right)}
\newcommand{\Htot}{H_{\text{tot}}}
\newcommand{\lr}[1]{\left(#1\right)}
\usepackage{simpler-wick}
\title{Deep Neural Nets as Hamiltonians}
\author{Mike Winer\footnote{University of Maryland, Institute for Advanced Study, mikewins@ias.edu}, Boris Hanin\footnote{Department of Operations Research and Financial Engineering, Princeton University, bhanin@princeton.edu} }

\begin{document}
\maketitle
\begin{abstract}
Neural networks are complex functions of both their inputs and parameters. Much prior work in deep learning theory analyzes the distribution of network outputs at a fixed a set of inputs (e.g. a training dataset) over random initializations of the network parameters. The purpose of this article is to consider the opposite situation: we view a randomly initialized Multi-Layer Perceptron (MLP) as a Hamiltonian over its inputs. For typical realizations of the network parameters, we study the properties of the energy landscape induced by this Hamiltonian, focusing on the structure of near-global minimum in the limit of infinite width. Specifically, we use the replica trick to perform an exact analytic calculation giving the entropy (log volume of space) at a given energy. We further derive saddle point equations that describe the overlaps between inputs sampled iid from the Gibbs distribution induced by the random MLP. For linear activations we solve these saddle point equations exactly. But we also solve them numerically for a variety of depths and activation functions, including $\tanh, \sin, \text{ReLU}$, and shaped non-linearities. We find even at infinite width a rich range of behaviors. For some non-linearities, such as $\sin$, for instance, we find that the landscapes of random MLPs exhibit full replica symmetry breaking, while shallow $\tanh$ and ReLU networks or  deep shaped MLPs are instead replica symmetric. 
\end{abstract}

\section{Introduction}
\label{sec:Intro}

The function $H(x;\theta)$ computed by a neural network has two kinds of variables: inputs $x$ and parameters $\theta$. An important chapter in deep learning theory is the analysis of the distribution of networks outputs $H(x;\theta)$ when each network parameter is chosen independently at random at the start of training \cite{neal1996priors,lee2017deep,Roberts_2022,poole2016exponential,yaida2020nongaussianprocessesneuralnetworks,matthews2018gaussian,li2023neuralcovariancesdeshaped,hanin2021randomneuralnetworksinfinite,hanin2018neural,hanin2018start,hanin2020products,hanin2023randomfullyconnectedneural}. This is typically done for one of two reasons:
\begin{itemize}
    \item[(i)] \textbf{Hyperparameter selection in model scaling.} The goal here is to determine how to choose network initialization (weight/bias variances), parameterization (normalizations of various layers and network components), and optimization hyperparameters (learning rate, weight decay) to obtain a neural network with numerically stable forward and backward passes, independent of model width, depth, input dimension/structure, and so on. The core idea is to analyze the initial values and first step updates to various order parameters given by averages over neurons in each hidden layer. Requiring that these order parameters and their updates remain order $1$ as model scale diverges gives a set of principles for setting initialization, parameterization, and optimization hyperparameters \cite{bordelon2023depthwise,bordelon2024infinite,yang2021tensor, yaida2022meta, yang2023tensor}.
    \item[(ii)] \textbf{Statistical mechanics of learning with neural networks.} The goal here is to analyze the Gibbs distribution (i.e. Bayesian predictive posterior) over network outputs for the Hamiltonian $-\log \mathbb P_{\text{iid}} - \log \mathcal L_{\text{data}}$, where $\mathbb P_{\text{iid}}$ is the distribution over network outputs when each network parameter is sampled independently and $\mathcal L_{\text{data}}$ is an empirical risk over a given training dataset. Since $\mathcal L_{\text{data}}$ is typically simple as a function of network outputs, the main technical difficulty is to obtain a good understanding of the distribution of network outputs under $\mathbb P_{\text{iid}}$ \cite{hanin2023bayesian,hanin2024bayesianinferencedeepweakly,ingrosso2024statistical,pacelli2023statistical,bassetti2024feature}.
\end{itemize}
\begin{figure}
    \centering
    \includegraphics[width=0.45\linewidth]{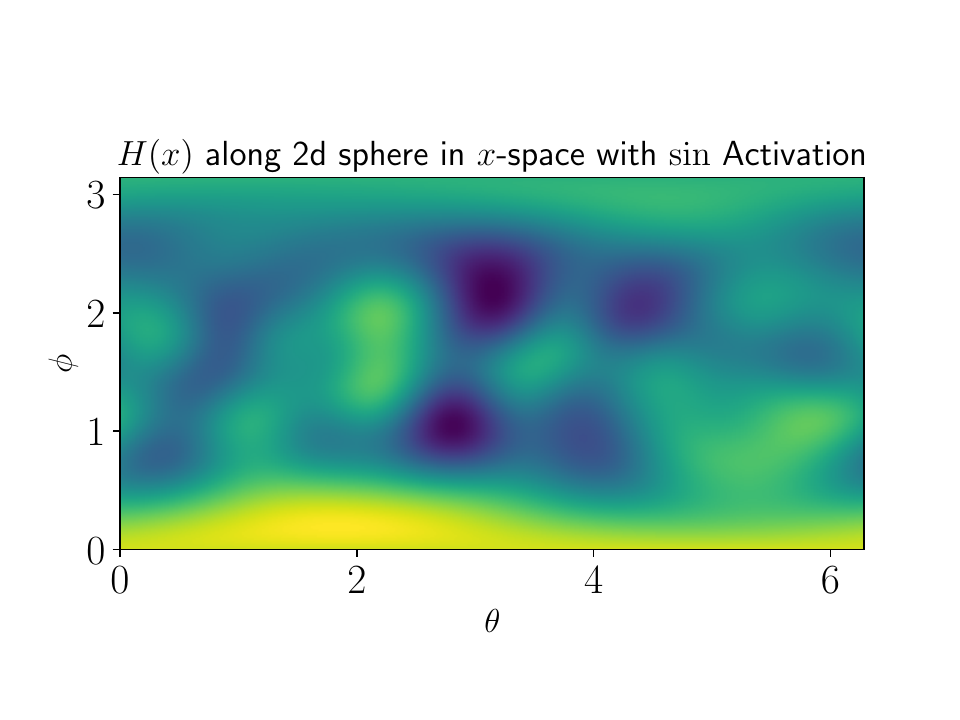}
    \includegraphics[width=0.45\linewidth]{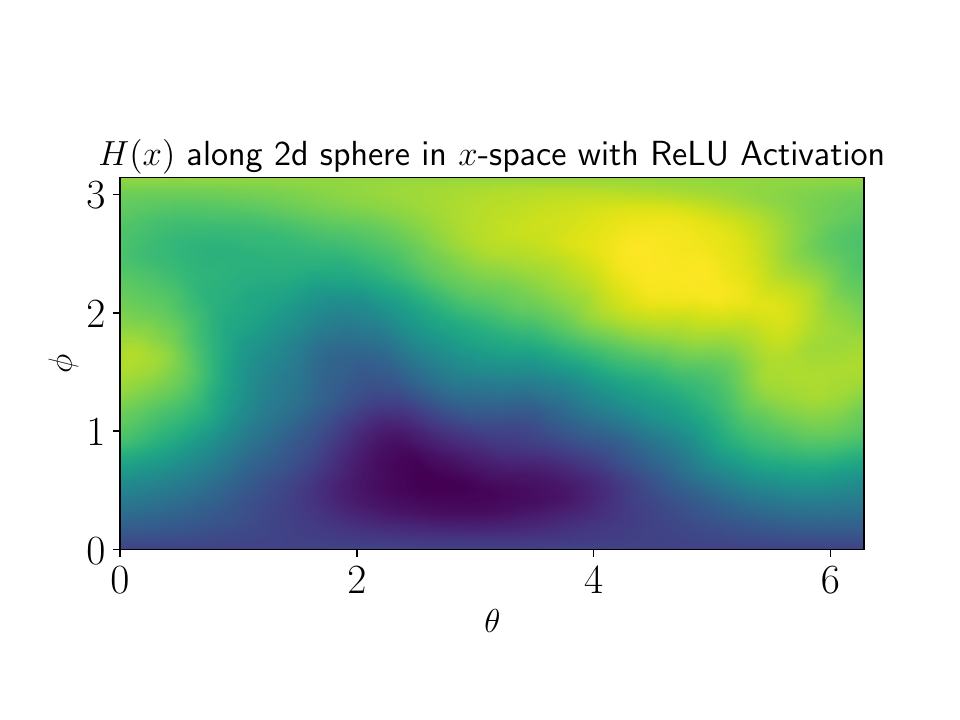}
    \caption{Outputs for a random 1-hidden-layer neural network for two different activation functions. In all cases, the input is a 2-sphere in the input space, with radius scaling as $\sqrt N$. For $\sin$ activations, we see a complicated function with many similarly-deep local minima. For ReLU activation, we find only a single local minimum on the  sphere. 
    }
    \label{fig:landscapes}
\end{figure}
In both (i) and (ii), one typically views the training data as quenched and analyzes the distribution over network outputs by annealing -- i.e. computing expectations -- over the parameter initialization. The purpose of this article is to understand the opposite situation in which network parameters are quenched while network inputs are annealed. More precisely, for the simple but fundamental case of fully connected networks $H(x;\theta)$, we will consider a disordered Hamiltonian 
\begin{equation}\label{eq:H-def}
\Htot(x;\theta) =\frac{1}{2\beta}\norm{x}^2  + H(x;\theta),    
\end{equation}
in the input variable $x$ with quenched disorder $\theta$ drawn from an distribution with independent centered components (see \S \ref{subsec:mlp-def} for a precise definition). Our goal is to determine the structure of the Gibbs measure $\exp(-\beta \Htot)$ in the setting where the network width (and sometimes depth) tends to infinity. Since $\exp(-\beta \Htot)$ concentrates around network inputs $x$ for which $H(x;\theta)$ is large and negative, we are really asking the following 
\begin{align}
   \label{eq:Q} \text{\underline{Question}:\quad }&\text{\textit{What is the distribution of network inputs with unusually large (negative) output values?}}
\end{align}
 We are motivated to study this question for several reasons. The first is its relation to adversarial perturbations \cite{goodfellow2014explaining}. A well-known and ubiquitous property of \textit{trained} neural networks is the presence of so-called adversarial examples. By definition, an adversarial example is a perturbation $x+ \delta x$ of a network input $x$ -- often an example from the training or validation set -- at which the network prediction at $x+\delta x$ differs radically from its value at $x$, despite $\delta x$ being small in some absolute sense. While practitioners are most interested in adversarial examples after training, there have been several prior theoretical analyses which indicate that for fully connected ReLU networks, adversarial examples exist with high probability \textit{at initialization}. Like these analyses, we also study the behavior of $H(x;\theta)$ for a typical value of $\theta$ \cite{bubeck2021single,bartlett2021adversarial}. However, unlike these works, which consider fluctuations of $H(x;\theta)$ that scale like a large constant times $H(x;\theta)$, we are more interested in the \textit{extreme} values of $H(x;\theta)$, which as we'll see are usually $\sqrt{\text{width}}$ larger than the the typical value of $H(x;\theta)$. For activations such as $\mathrm{ReLU}$  (right panel of Figure \ref{fig:landscapes}) or $\tanh$ (Figure \ref{fig:landscapes-tanh}) we will see  that all thermodynamically relevant minimum of $H(x;\theta)$ tightly clustered at large width. In contrast, for some activations such as $\sin$ there  may be many well-separated and thermodynamically relevant minima (left panel of Figure \ref{fig:landscapes}).

The second reason we are interested in Question \eqref{eq:Q} above is related to feature visualization. The context for this is that neural networks are remarkably effective algorithms at feature learning, i.e. the process of learning useful non-linear transformations of the training data. This underlies the common pipeline of pre-training and then fine-tuning, for instance, in which a large neural network is first trained on an upstream dataset, and then some part of the neural network, perhaps the encoder in an LLM or the mapping from input to the final hidden layer representation in a convolutional network, is reused to extract features for solving downstream tasks on which there may be much less training data. Motivated in part by the practical success of such approaches, a rich line of empirical work has sought to understand or explain what kinds of features a given neural network has learned \cite{yosinski2014transferable,yosinski2015understanding}. In particular, to interpret the features learned by a given neuron (or group of neurons), it is common to apply \textit{feature maximization}. That is, for a fixed setting of learned parameters $\theta$, one seeks to extremize the value of the pre-activation of this neuron as a function of the network input. From this point of view, the present article analyzes a kind of null model in which we extremize the pre-activation of a given neuron (we happen to choose the network output but this is immaterial at initialization) as a function of its input. We believe the techniques developed here will be useful in analyzing the nature of learned features after training by Bayesian inference as well, and we plan to return to this in a future article. 

\begin{figure}
    \centering
    \includegraphics[width=0.4\linewidth]{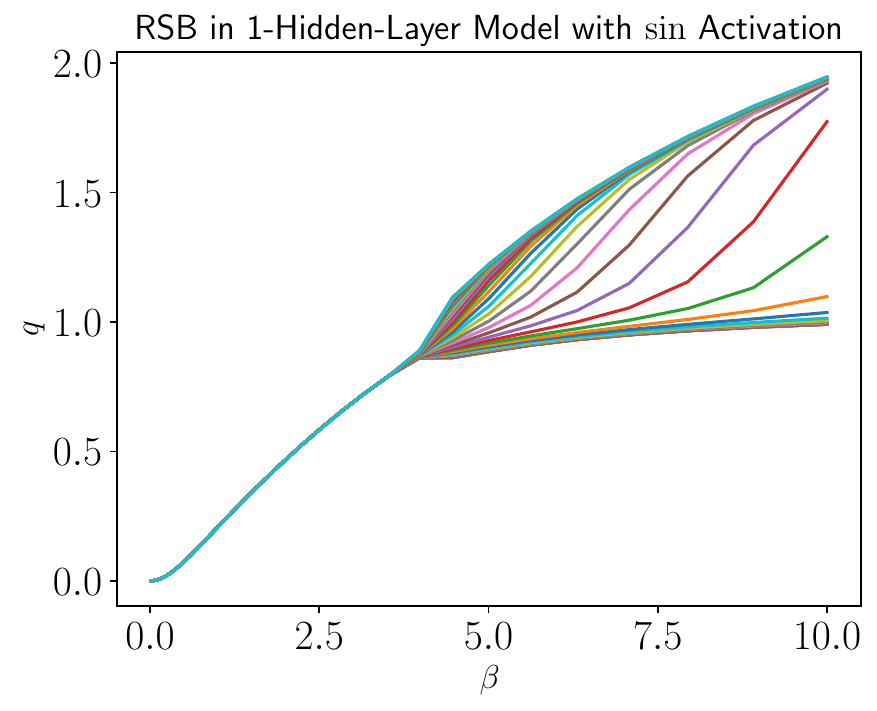}
    \includegraphics[width=0.4\linewidth]{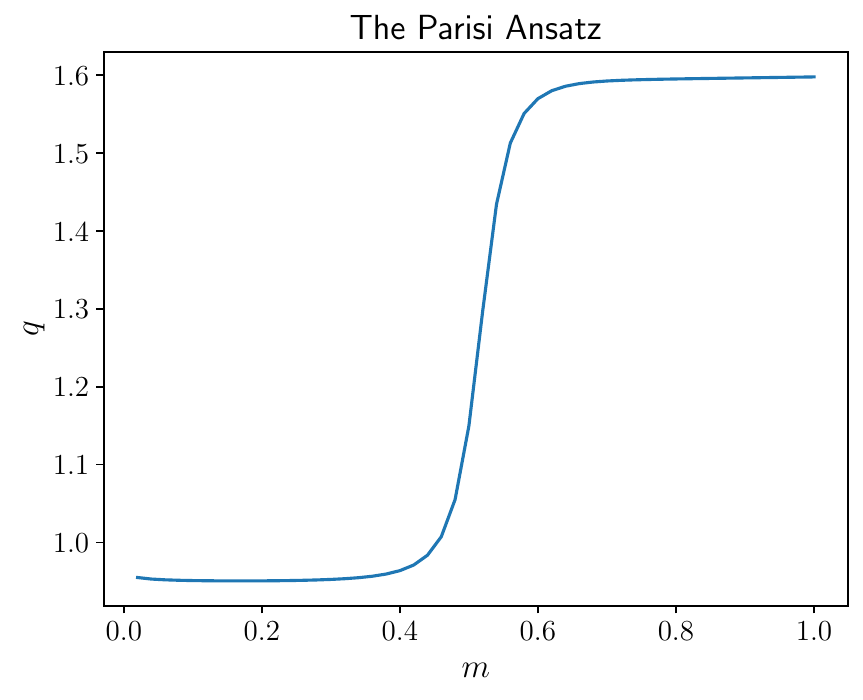}
    \caption{We study the distribution of overlaps $q$ between two inputs draws independently from the Gibbs distribution $\exp(-\beta \Htot)$ coming from a one layer network with $\sin$ activations at infinite width. (Left) Graph of $q$ as a function of $\beta$ for a 50-step RSB ansatz. For $\beta$ below approximately $4.5$, every $q$ is the same because the system is replica symmetric. At larger $\beta$, however, the system exhibits what appears to be full RSB, with a standard bimodal $q$ distribution. (Right) Graph of $q$ as a function of $m$ for fixed $\beta=7.1$. It displays the standard Parisi behavior of having two constant sections connected by a non-constant piece.}
    \label{fig:RSBGraph}
\end{figure}

Our final motivation for studying Question \eqref{eq:Q} is as a proof of concept for using the replica method to analyze neural networks with quenched weights. This complements a number of prior articles that apply replica computations to in the context of annealed weights (and quenched inputs) \cite{cui2024asymptotics,cui2025high,seung1992statistical,gabrie2023neural}. To conclude the introduction, we give provide a brief and informal summary of our main results:
\begin{itemize}
    \item Let us denote by $Z_\beta$ be the partition function at inverse temperature $\beta$ for the Gibbs measure $\exp(-\beta \Htot)$ of the Hamiltonian \eqref{eq:H-def}. For every $n\geq 0$, we obtain saddle point equations for the action of the annealed $n$-replica partition function $\mathbb E[Z_\beta^n]$ in the limit as the network width tends to infinity. The variables in this action are $n$-replica overlap matrices $Q^{(\ell)}$ and certain dual variables $\Sigma^{(\ell)}$, where $\ell\geq 0.$ See \S \ref{subsec:saddlePoint} for the precise result. 
    
    \item For any fixed depth $L$ and non-linearity, we propose a simple numerical algorithm, which we call the \textit{zipper method}, for obtaining saddle points for the action of $\mathbb E[Z_\beta^n]$ in the limit as $n\rightarrow 0$. This allows us to study the disorder averaged log partition function $\mathbb E[\log Z_\beta]$ and hence the statistics of the Gibbs distribution $\exp(-\beta \Htot)$. We use the zipper method to provide numerical evidence for the following statements:

    \begin{itemize}
        \item For one layer networks with $\sin$ activations, the Gibbs measure $\exp(-\beta \Htot)$ exhibits full RSB (see Figure \ref{fig:RSBGraph}). This suggests that for a typical setting of network weights $\theta$ the set of inputs $x$ for which the Gibbs measure $\exp(-\beta \Htot(x;\theta))$ is spread over a small number of distinct clouds around local minima of $\Htot$. These minima are embedded in high-dimensional space, and obey the ultrametric triangle inequality $d(a,b)\leq \max(d(a,c),d(b,c))$.
        \item We also analyze other simple architectures with common activations including ReLU and tanh. We find replica symmetry for all temperatures we investigate. This does not, however, guarantee that the function has only one local minimum, only that only one region of phase space is thermodynamically important. $\tanh$ activations, for instance, seem to lead to functions with many local minima but replica-symmetric thermodynamics (see Figure \ref{fig:landscapes-tanh}). In principle, these minima can be counted using replica-based tools like the Franz-Parisi Potential \cite{Franz_1995,Monasson_1995}. 
    \end{itemize}
    \begin{figure}
    \centering
    \includegraphics[width=0.45\linewidth]{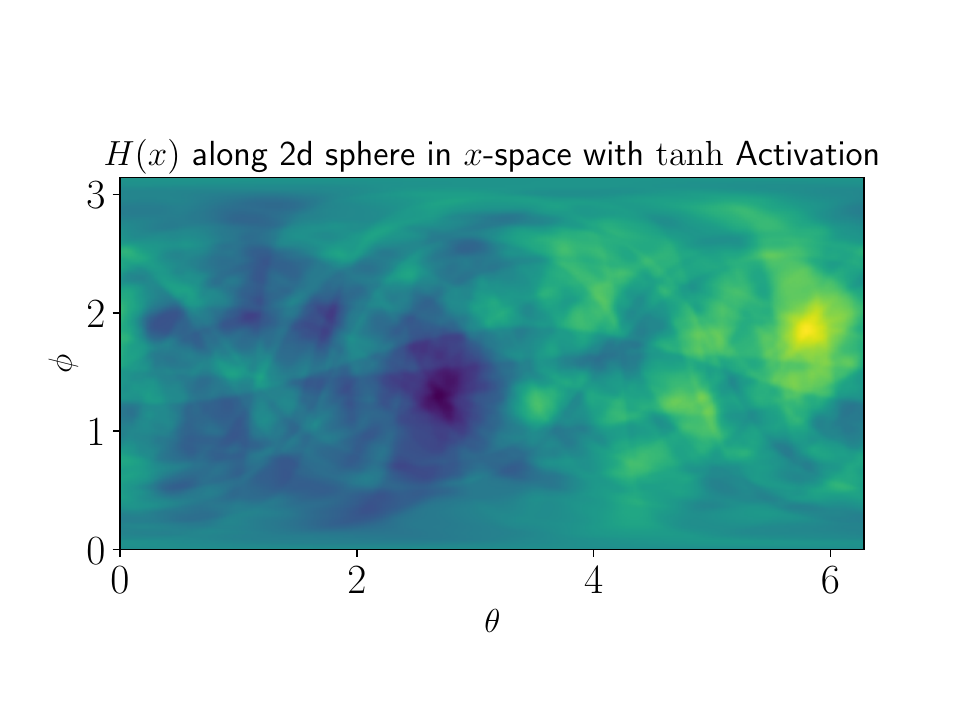}
    \caption{Outputs for a random 1-hidden-layer neural network with $\tanh$ activations. The input is a 2-sphere in activation space, with radius scaling as $\sqrt N$. We see multiple local minima, but one is clearly dominant.}
    \label{fig:landscapes-tanh}
\end{figure}
    \item We further investigate analytically the disorder averaged log partition function $\mathbb E[\log Z_\beta]$ for odd shaped non-linearities of the form
    \[
        \sigma(t) = t+\frac{\psi}{3L}t^3,\qquad \psi \in \mathbb R
    \]
    in the setting where first the network width $N$ and then the network depth $L$ diverge.\footnote{Such shaped non-linearities, proposed originally in \cite{li2023neuralcovariancesdeshaped,martens2021rapid}, can be thought of as rescalings $\sigma(t)\approx \sqrt{L}\phi(t/\sqrt{L})$ for an odd function $\phi$. The original motivation for studying these non-linearities is that power $\alpha = 1$ in the $L^{-\alpha}$ pre-factor in front of the $t^3$ term is the smallest value of $\alpha$ for which the two point function remains non-degenerate at large depth at infinite temperature (see also \cite{Roberts_2022, hanin2024bayesianinferencedeepweakly}).} We find

    \begin{itemize}
        \item For deep linear networks ($\psi=0$), we give exact solutions to the saddle point equations for disorder averaged log partition function $\mathbb E[\log Z_\beta]$. We see that the overlap matrices at the final hidden layer are give by constant times all ones matrix unless $\beta^2 L =O(1)$. The Gibbs measure $\exp(-\beta \Htot)$ is dominated by replica symmetric overlap matrices since the Hamiltonian is quadratic. See \S \ref{sec:DLN}.
        \item For general $\psi$, in the high temperature regime $\beta^2 L = O(1)$ we reduce extremizing the action for the disorder averaged log partition function $\mathbb E[\log Z_\beta]$ to a solving a system of two coupled non-linear ODEs. Numerically solving these ODEs suggests that deep shaped networks give a replica symmetric overlap structure. See \S \ref{sec:shaped}. 
        \item In part to validate the correctness of both our saddle points and the zipper method, we check that  for larger values of $\beta$ correspond to overlap matrices $Q^{(\ell)}$ with off-diagonal entries closer to one dominate the disorder averaged log partition function (see Figure \ref{fig:qsTogether}). 
    \end{itemize}
    
\end{itemize}

\section{Related Literature}


The celebrated Neural Network-Gaussian Process (NNGP) correspondence \cite{neal1996priors,lee2017deep,hanin2021randomneuralnetworksinfinite,g.2018gaussian,garrigaalonso2019deepconvolutionalnetworksshallow} is the observation that at large hidden-layer width, the output after $\ell$ layers of a random neural network is a centered Gaussian Process with $2$-point function $K^{(\ell)}$, which satisfies a simple recursion with respect to $\ell$. Unsurprisingly, in the infinite temperature regime $\beta=0$, the saddle points equations for the annealed $n-$replica partition function $\E{Z_0^n}$ are precisely given by $K^{(\ell)}$ evaluated at all pair of replicas. Hence, our computations can be seen as a generalization of those recursions to extreme elements, just as extreme value theory is a generalization of the central limit theorem to extreme events. 

While we do not pursue the computation of finite width corrections in this article, there is in principle no obstruction to doing so, and we expect to obtain generalization of prior work studying the non-Gaussianities for fixed inputs beyond the infinite-width limit \cite{yaida2020nongaussianprocessesneuralnetworks,Roberts_2022,hanin2023randomfullyconnectedneural}. In this vein, we also point the reader to \cite{macci2024largemoderatedeviationsgaussian} looks not at perturbative corrections to Gaussianity, but at extreme deviations of the output for fixed input with a varying neural net.

Our paper can be thought of turning these questions on their head in the simplest case of infinite width, finite depth networks before training. Rather than probing the distribution for fixed inputs and a random neural network, we probe the distribution for fixed (typical) neural networks and random inputs.

Another set of papers relevant to our work  analogizes neural networks to energy landscapes, but treats the training loss as the energy, with the weights being the microscopic variables \cite{choromanska2015losssurfacesmultilayernetworks,pmlr-v40-Choromanska15,liao2017theoryiilandscapeempirical,liao2024exploringlosslandscapeslens}. This approach is responsible for tremendous insights, including explaining the power and failures of SGD to minimize losses.

There is also a large body of work in the physics literature studying Gaussian landscapes, and the associated thermodynamics. This includes work for generic Gaussian landscapes \cite{Bray_2007,Seyed_allaei_2008}, and landscapes induced on the sphere \cite{Crisanti_2003,Crisanti_2006,barrat1997pspinsphericalspinglass,Bates_2022,zhou2023sphericalmixedpspinglass}. The article \cite{lacroixacheztoine2024superpositionplanewaveshigh} goes further, and studies the landscapes generated by what we could call a one-hidden-layer perceptron with all positive weights.

We conclude our review of literature by highlighting several strands of prior work that compute saddle points for the action of various versions of the annealed partition functions $\mathbb E(Z_\beta^n)$ for quenched network inputs. The first line of work is related to Bayesian inference and includes the articles of Cui et. al. \cite{cui2024asymptotics,cui2025high}, which rely on the replica method and a range of work \cite{hanin2023bayesian,hanin2024bayesianinferencedeepweakly, ingrosso2024statistical,pacelli2023statistical, bassetti2024feature,li2021statistical, aiudi2025local, seroussi2022separationscalesthermodynamicdescription,naveh2021self,fischer2024critical} which focuses on  statistical mechanics (e.g. Bayesian inference) of wide neural networks and specifically on the distribution of overlap matrices under both the prior and posterior. The second line of work is the DMFT analysis of Bordelon et. al. \cite{bordelon2023depthwise,bordelon2024infinite, bordelon2022self}, which obtains not only saddle point equation for the overlap matrices $Q^{(\ell)}$ but also their analogs in the backward pass and the dynamics of these overlaps during training.

\subsection*{Plan For Remainder of This Paper}

The remainder of this paper is organized as follows. Section \ref{sec:replica} starts by defining the MLPs we study and proceeds by deriving and analyzing the resulting replicated action. It opens with an overview of the replica trick as a way to study the thermodynamics of disordered systems. Specifically, subsection \ref{subsec:repAction} define a generalization of the overlap matrix for each layer of the MLP, and derives an action in terms of these matrices. Subsection \ref{subsec:saddlePoint} then derives the saddle-point equations for this action, and subsection \ref{subsec:n0} takes the number of replicas $n$ to 0.

Section \ref{sec:DLN} solves those equations in the case of linear activation functions, and compares this with a completely separate calculation of the overlap matrices using random matrix methods. It finds complete agreement in the large-$N$ limit.

Section \ref{sec:shaped} considers networks with \textit{shaped activations} \cite{li2023neuralcovariancesdeshaped,hanin2024bayesianinferencedeepweakly,martens2021rapid}. In the limit of growing depth (after first taking the width to infinity) we derive a differential equation for the overlap matrices as a function of depth, getting a result that generalizes the exact computations for deep linear networks from section \ref{sec:DLN}.

Section \ref{sec:Numerics} is the most numerically-focused portion of the paper. In it, we use Monte Carlo methods to sample from the Boltzmann distribution from MLPs with widths in the hundreds, comparing energy as a function of $\beta$ with the solution to the equations of motion. We find strong agreement between the two quantities. We then provide some final remarks in section \ref{sec:Conclusion}.

\section{The Replica Trick and Multilayer Perceptrons}
\label{sec:replica}
\subsection{Defining the Multilayer Perceptron}\label{subsec:mlp-def}
Our Hamiltonian $H$ is a complicated function of the inputs $x$. We will define
\begin{align}
    &z_i \l 1(x)\equiv \sum_{j=1}^{N\l 0}W\l 1_{ij}x_j\quad \text{for }1\leq i\leq N\l 1\\
    &z_i \l {\ell+1}(x)\equiv \sum_{j=1}^{N\l{\ell}}W\l {\ell+1}_{ij}\sigma \left(z \l \ell_j\right)\quad \text{for }1\leq \ell\leq L-1,1\leq i\leq N\l {\ell+1}\\
    &H(x)\equiv \sum_{i=1}^{N\l L}W\l {L+1}_i \sigma(z\l L_i) 
\end{align}
In words, our function is calculated in terms of \textit{layers}. The $\ell$th layer has width $N\l \ell$, and is calculated in terms of the previous layer. We will take the limits where all of the $N^{(\ell)}$ are proportional to a single large parameter $N$, which goes to infinity 
\[
N^{(\ell)} = a^{(\ell)} N,\quad \ell=1,\ldots, L,\quad N\gives \infty.
\]
We will call $\sigma(z\l \ell_i)$ the activation of the $i$th neuron in layer $\ell$, and $z\l \ell_i$ the preactivation. Note that $W\l \ell$ is an $N\l{\ell}\times N\l {\ell-1}$ matrix. We take its entries to be iid centered Gaussians with variance $1/N\l{\ell-1}$, except for $W\l {L+1}$ whose entries have variance $1.$ This convention is chosen so that the scale of pre-activations $z_i^{(\ell)}$ is independent of $N$, but the overall Hamiltonian will be extensive in $N$. 

\subsection{Warmup: Single System Partition Function and Three Kinds of Averages}

By definition, for a given realization of the network weights $W_{ij}^{(\ell)},$ the partition function of a single system is
\begin{equation}
    Z_\beta=\int \exp(-\beta \Htot(x)) \frac{d^{N\l 0} x}{\sqrt{2\pi}^{N\l 0}}=\int  \exp(-\beta H(x)) \exp\left(-\frac 12 \norm{x}^2\right) \frac{d^{N\l 0} x}{\sqrt{2\pi}^{N\l 0}}.
\end{equation}
This partition function can also be thought of as the expected value of $\exp(-\beta H(x))$ when $x$ is drawn from a multivariate Gaussian. Since we are selecting a random neural net with random weights, $Z$ is a random variable. At this point, it is worth making clear the three types of average we are juggling:
\begin{itemize}
    \item \textbf{Disorder Averages}: There are averages over the weights of the neural nets, which will be denoted by $\mathbb E$. It is in this context that it makes sense to talk about the average partition function of a given architecture, average free energy, or the Kernel (covariance between outputs of two fixed inputs). The, terminology disorder average is borrowed from condensed matter physics where random variations in physical samples results in random Hamiltonians.
    \item \textbf{Thermodynamic Averages}: These are averages over the thermodynamic ensemble, i.e. with respect to the Boltzmann distribution
    \[
    p(x)\sim \frac{1}{\sqrt{2\pi}^{N\l 0}}\exp\left(-\frac 12 \norm{x}^2-\beta H(x)\right).
    \]
    In this sense, it makes sense to talk about the average value of $x$ or $H(x)$ for a fixed neural net. These two types of averages can also be combined, one can take $\mathbb E\left(\int dx p(x) \bullet\right)$. This is the disorder average of the thermodynamic average of a quantity.
    \item \textbf{Gaussian Averages}: These are averages over a Gaussian input distribution 
    \[
    \frac{1}{\sqrt{2\pi}^{N\l 0}}\exp\left(-\frac 12 \norm{x}^2\right). 
    \]
    These averages are just mathematical tricks, we can make use of the fact that the partition function for normalizing thermodynamic averages is the average of $e^{-\beta H(x)}$ under this type of average. We will sometimes combine the Gaussian and Disorder average, since the Partition function is the disorder average of the Gaussian average of $e^{-\beta H}$.
\end{itemize}

We will be interested in $\mathbb E(Z_\beta^n)$, the value of the \textit{$n$ replica partition function} $Z_\beta$ averaged over the microscopic weights. Even more important to us, however, will be $\mathbb E(\log Z_\beta)$ because if we are interested in something like the disorder average of the thermodynamic average of the energy $H(x)$, this can be expressed as
\begin{equation}
    \text{disorder average of }H(x)=-\frac{\partial}{\partial \beta}\mathbb E(\log Z_\beta)\propto \mathbb E \left(\int  H(x)\exp(-\beta H(x)) \exp\left(-\frac 12 \norm{x}^2\right) d^{N\l 0} x\right).
\end{equation}

Other similar observables, besides energy, are also best understood in terms of $\mathbb E(\log Z_\beta)$. In some areas, one can make the approximation $\mathbb E (\log Z)\approx \log \mathbb E (Z)$. However, as we will see, this approximation is not even close for the Hamiltonians studied in this work. As we explain in the next section this is true even in the very simplest Hamiltonians we consider. In order to evaluate $\mathbb E (\log Z_\beta)$ we rely on a method  called the \textbf{replica trick}, which we introduce next. 

\subsection{The Replica Trick For Rookies}
\label{subsec:toymodel}
The replica trick relies on two key insights, both simple in their way. The first is that
\begin{equation}\label{eq:replica-trick-1}
    \mathbb E(\log Z_\beta)=\lim_{n\to 0} \partial_n\mathbb E(Z_\beta^n)=\lim_{n\to 0}\frac 1n \log \E{Z_\beta^n}.
\end{equation}
The second is that for positive integer $n$
\begin{equation}
        \mathbb E\left(Z_\beta^n\right)=\int  \mathbb E\left(\exp\left (-\beta \sum_{a=1}^n H(x_a )\right)\right) \exp\left(-\frac 12 \sum_{a=1}^nx_a^2\right) \frac{d^{nN\l 0} x_a }{\sqrt{2\pi}^{nN\l 0}}.
        \label{eq:repPartition}
\end{equation}

The $n$ independent vectors $x_a$, each independent of the others conditioned on the value of the network parameters, are called the $n$ replicas. In words, this equation is saying that the $n$th power of the partition function of a supersystem consisting of $n$ copies, or replicas, of our original neural net acting on $x$. The surprising insight of the replica trick is that one can find a formula for equation \ref{eq:repPartition} directly computing saddle points equations for the effective action in \eqref{eq:repPartition} and then analytically continuing $n\to 0$. 

The rest of this section will be spent working through an extremely simple example so as to illustrate two key ideas: the usefulness of \textit{overlap matrices}, and the power of the \textit{saddle-point approximation}. The Hamiltonian we will consider in this section is simply
\begin{equation}
    H(x)=\sum_{i=1}^N h_i x_i,
\end{equation}
where the elements of $h$ are drawn iid from the standard normal distribution. This is technically a special case of the multilayer perceptron Hamiltonian discussed earlier, with no hidden layers and a linear activation function. The partition function of this system is then
\begin{equation}
    Z_\beta=\int \exp\left(-\beta h\cdot x-\norm{x}^2/2\right)\frac{d^N x}{\sqrt{2\pi}^N}=\exp\lr{\beta^2 h^2/2}.
\end{equation}
Hence, 
\begin{equation}
    \log \mathbb E (Z_\beta)=-\frac{N}{2}\log \left(1-\beta^2\right),
    \label{eq:AnnealedToy}
\end{equation}
while
\begin{equation}
    \mathbb E(\log Z)=N\beta^2/2.
    \label{eq:QuenchedToy}
\end{equation}
These two quantities are obviously different from each other! On the other hand, let us fix $n\geq 1$ and compute the replicated partition function:
\begin{equation*}
    Z_\beta^n=\int \exp\left(-\sum_{a=1}^n \beta h\cdot x-x^2/2\right)\frac{d^{nN} x}{\sqrt{2\pi}^{nN}}
\end{equation*}
For this simple system, we can of course calculate $\mathbb E(Z_\beta^n)$ directly for arbitrary $h$, but for illustrative purposes we will take the $h$ integral first. This gives 
\begin{equation}\label{eq:Zn-simple-with-Q}
    \mathbb E(Z_\beta^n)=\int \exp\left(-\frac{N\beta^2}{2}\sum_{a,b=1}^N \frac 1N x_a\cdot x_b-\sum_{a=1}^nx^2/2\right)\frac{d^{nN} x}{\sqrt{2\pi}^{nN}}
\end{equation}
By integrating out $h$, we have correlated the different replicas. This is reflected in the appearance of the matrix $Q_{ab}=\frac{1}{N} x_a\cdot x_b$, which is called the \textit{overlap matrix} and is a measure of how similarity between the configurations of replicas $a$ and $b$. To further simplify the expression in \eqref{eq:Zn-simple-with-Q} we will make use of the standard fact that
\begin{equation}
    f(u)=\frac{2\pi }{N}\int D\Sigma DQ \exp(N\Sigma(Q-u)) f(Q)
\end{equation}
for an essentially arbitrary function $f$, where the contour for the $\Sigma$ integral is parallel to the imaginary axis. In terms of the saddle point approximation we will perform later, this actually means we will be minimizing with respect to $\Sigma$, not maximizing. This allows us to write
\begin{equation}
    \mathbb E(Z_\beta^n)= \int D\Sigma DQ \exp\left(-\frac{N\beta^2}{2}\sum_{a,b=1}^N Q_{ab}+\frac 12 \sum_{ab}\Sigma_{ab}\left(NQ_{ab}-x_a\cdot x_b\right)-\sum_{a=1}^n\norm{x_a}^2/2\right)\frac{d^{nN} x}{\sqrt{2\pi}^{nN}},
\end{equation}
where we drop some pesky factors for $\sqrt{2\pi/N}^{n^2}$ which will go to 1 anyways. We will adopt the notation from physics literature, where $\Sigma$ is considered a \textit{self-energy}. However, we note that in neural network literature this quantity is sometimes written $\hat Q$, where it is considered a \textit{dual kernel}. At this point, the $x$ integral is purely Gaussian and we have

\begin{equation}
\begin{split}
    \mathbb E(Z_\beta^n)&= \int D\Sigma DQ \exp\left(S[\Sigma,Q]\right),\\
    S&=-\frac{N\beta^2}{2}\sum_{a,b=1}^N Q_{ab}+\frac N2 \sum_{ab}\Sigma_{ab}Q_{ab}-\frac N2 \log \det \left(\Sigma+\delta_{ab}\right).
\end{split}
    \label{eq:firstLargeN}
\end{equation}

Notice that the quantity in the exponent of \eqref{eq:firstLargeN} is proportional to $N$. For large $N$, we can use steepest descent (a generalization of Laplace's method) to argue that the integral should be approximated by $\exp\lr{S(\Sigma^*,Q^*)}$, where $\Sigma^*,Q^*$ is a stationary point of $S$. To find the stationary point we take derivatives to obtain
\begin{align*}
\frac{\partial S}{\partial \Sigma_{ab}}&= -\frac{N}{2}(\Sigma+\delta_{ab})^{-1}_{ab}+\frac N2 Q_{ab},\qquad  \frac{\partial S}{\partial Q_{ab}}=\frac N2 \Sigma_{ab}-\frac{N\beta^2}{2}.    
\end{align*}
Setting these both to zero we get
\begin{equation}
    Q_{ab}^*=\left(\delta_{ab}-\beta^2\right)^{-1}.
\end{equation}
In other words, we need to invert the matrix which is equal to the identity minus $\beta^2$ times the all $1$'s matrix. This matrix has one eigenvector which is proportional to $(1,1,1,...,1)$, whose eigenvalues is $1-n\beta^2$. The other $n-1$ eigenvalues are degenerate, and equal to 1. Therefore
\begin{equation}
    Q_{ab}^*=\delta_{ab}+\frac 1n \left(\frac{1}{1-n\beta^2}-1\right).
\end{equation}
Plugging this back into the action and using the steepest descent approximation gives
\begin{equation}
    \log \mathbb{E}(Z_\beta^n)=-\frac{nN\beta^2}{2}\frac{1}{1-n\beta^2}+\frac{nN}{2}-\frac{nN}{2}\left(1+\frac 1n\left(1-\frac{1}{1-n\beta^2}\right)\right)-\frac N2 \log (1-n\beta^2).
\end{equation}
This simplifies a quite bit to
\begin{equation}
    \log \mathbb{E}(Z^n)=-\frac N2 \log (1-n\beta^2).
\end{equation}
The key point is that this agrees with our $n=1$ calculation in  \eqref{eq:AnnealedToy} and also, by taking $n\gives 0$ in \eqref{eq:replica-trick-1} with the $n=0$ result in equation \eqref{eq:QuenchedToy}.

Our strategy for this simple problem was to rewrite our partition function in terms of an $n\times n$ overlap matrix $Q$ and Lagrange multiplier $\Sigma$, get a large action proportional to system size, solve the stationary point of this action, and then take the $n\to 0$ limit. It is this same basic strategy we will use to compute $\mathbb E(Z_\beta^n)$ and $\E{\log(Z_\beta)}$ for the multilayer perceptron below. 

\subsubsection{An Aside on Replica Symmetry}
\label{subsubsec:RS}
The action in \eqref{eq:firstLargeN} has an $S_n$ permutation symmetry, where switching any two replicas leaves the action unchanged. This is, in fact, a general feature of any system: when you have a collection of identical objects, switching their configurations won't change the energy or entropy. In the case of the toy model in this subsection, not only did the action and corresponding equations of motion respect this symmetry, the unique solution does as well. This is not always the case. Systems from the Sherrington-Kirkpatrick model \cite{parisiSK,Panchenko_2012} to hard spheres \cite{Parisi_Urbani_Zamponi_2020} to quantum optical systems \cite{kroeze2024directlyobservingreplicasymmetry} exhibit a spontaneous breaking of this symmetry. This goes by the name \textit{Replica Symmetry Breaking}, or RSB.

Although it may seem like a strange artifact of mathematical trickery, RSB actually has a clear physical interpretation: for a complicated energy function, there are several competing scenarios for the nature of the local minima. There maybe a unique local minimum, as one can manifestly see in the toy problem from the previous section. There may also be exponentially many local minima, with the the Boltzmann measure spread out between them. Or there may be many local minima, but the Boltzmann measure is condensed in just a few. In the first scenario, every pair of replicas is in the same minima. In the second, every pair will be in different minima. It is this last scenario that leads to RSB. Two replicas have a noticeable probability of inhabiting the same minima, or of different minima, thus different overlaps can be different. We will see later in this paper that certain activation functions can lead to the rich phenomenology of RSB (see Figure \ref{fig:RSBGraph}).

\subsection{The Replicated Action}
\label{subsec:repAction}

In this section, we will derive a path integral expression (i.e an explicit action) for the average value of $Z_\beta^n$:
\begin{equation}\label{eq:Zbeta-n}
        \mathbb E\left(Z_\beta^n\right)=\int  \mathbb E\left(\exp\left (-\beta \sum_{a=1}^n H(x_a )\right)\right) \exp\left(-\frac 12 \sum_{a=1}^n \norm{x_a}^2\right) \frac{d^{nN\l 0} x_a }{\sqrt{2\pi}^{nN\l 0}}.
\end{equation}
To do this, we define an $n\times n$ overlap matrix $Q^{(\ell)}$ in each layer:
\begin{align*}
    Q_{ab}\l 0&=\frac 1 {N\l 0}\sum_{i=1}^{N\l0} x_{ia}x_{ib},\qquad Q\l\ell_{ab}=\frac 1 {N\l \ell}\sum_{i=1}^{N\l\ell} \sigma(z\l \ell_{ia})\sigma(z\l \ell_{ib})\quad \text{for }1\leq \ell\leq L
\end{align*}
Notice that the integral in \eqref{eq:Zbeta-n} requires us to compute the Laplace transform of replica summed Hamiltonian $\sum_a H(x_a)$. Our strategy for performing this computation starts by noticing that while $\sum_a H(x_a)$ is a complicated function of $x_a$, it is in fact a simple function of the overlap matrix $Q^{(L)}$. We will then need to compute the distribution of $Q^{(L)}$ as a function of $x_a$. This is again a complicated distribution. However, just as in the toy model from \S \ref{subsec:toymodel}, will introduce dual  overlap matrices $\Sigma^{(\ell)}$ to efficiently compute the distribution of $Q{(\ell)}$ given $Q^{(\ell-1)}$ for all $\ell=L,\ldots, 1$. Finally, the distribution of $Q^{(1)}$ given $x_a$ is simple and we will obtain an explicit action for $\mathbb E(Z_\beta^n)$. 

To implement this plan, the key insight is that conditional on a given $Q^{(\ell)}_{ab}$, the preactivations of the next layer $z\l{\ell+1}_{ia}$ are jointly Gaussian with mean 0 and covariance
\begin{equation*}
    \textrm{Cov}\left(z\l{\ell+1}_{ia},z\l{\ell+1}_{jb}\right)=\delta_{ij}Q\l \ell_{ab}.
\end{equation*}
Hence, 
\begin{equation*}
\mathbb P(Q^{(\ell+1)}|Q^{(\ell)})=\int D^{N\l{\ell+1}}z^{(\ell+1)}D\Sigma^{(\ell+1)}_{ab}\exp\left(-\frac 12\sum_{ab}\Sigma^{(\ell+1)}_{ab}\left(N\l{\ell+1}Q^{(\ell+1)}_{ab}-\sum_i\sigma(z^{(\ell+1)}_{ia})\sigma(z^{(\ell+1)}_{ib})\right)\right),
\end{equation*}
where the integral is over the space of Lagrange multiplier $n\times n$ matrices $\Sigma^{(\ell+1)}_{ab}$ and the measure $D^{N\l{\ell+1}}z\l {\ell+1}$ is shorthand for the Gaussian integrand
\begin{equation*}
    D^{N\l{\ell+1}}z\l{\ell+1}=\frac{d^{nN\l{\ell+1}}z\l {\ell+1}_{ia}}{\sqrt{\det 2\pi Q\l \ell}^{N\l \ell}}\exp\left(-\frac 12 \sum_i \sum_{ab}z\l {\ell+1}_{ia} \left(Q\l \ell_{ab}\right)^{-1}z\l {\ell+1}_{ib}\right).
\end{equation*}
To simply the notation, let us define 
\begin{align}
    \notag \Zsq{\Sigma_{ab}}{Q_{ab}}&\equiv \int \exp\left(-\frac 12 \sum_{ab} z_a \left(Q_{ab}\right)^{-1} z_{b}+\frac 12 \Sigma_{ab}\sigma(z_a)\sigma(z_b)\right)\frac{d^n z}{\sqrt {\det 2\pi Q }}\\
  \label{eq:ZsqDef}&=\int  \exp\left(\frac 12 \sum_{ab}\Sigma_{ab}\sigma(z_a)\sigma(z_b)\right) Dz
\end{align}
to be the expected value of $\exp \left(\frac 12\Sigma_{ab}\sigma(z_a)\sigma(z_b)\right)$ where $z$ is a Gaussian with covariance $Q$. Now we can write
\begin{equation*}
\mathbb P(Q^{(\ell+1)}|Q^{(\ell)})=\int D\Sigma^{(\ell+1)}_{ab}\exp\left(-\frac 12N\l{\ell+1}\sum_{ab}\Sigma^{(\ell+1)}_{ab}Q^{(\ell+1)}_{ab}
+N\l {\ell+1}\log \Zsq{\Sigma \l {\ell+1}}{Q \l \ell}\right).
\end{equation*}
Just as we derived the distribution of $Q\l {\ell+1}$ conditioned on $Q\l \ell$, we can also derive the distribution of $H$ conditioned on $Q\l L$. If we condition on a given value of $Q\l L_{a b}$, the $n$ variables $H(x_{ia})$ are joint Gaussian with covariance $N\l L Q\l L_{a b}$. We thus have
\begin{equation*}
    \mathbb E \left(Z^n(b)|Q\l L_{ab}\right)=\mathbb E \left(\exp\left(-b\sum_{a} H(x_{ia}\right)|Q\l L_{ab}\right)=\exp\left(-\frac 12 \beta^2N\l L \sum_{ab} Q\l L_{ab}\right)
\end{equation*}

We now have a way to get the distribution of $H$ conditioned on $Q\l L$ and the distribution of $Q\l{\ell+1}$ conditioned on $Q\l \ell$. All we need is an expression for the probability distribution of $Q\l 0$ in the Gaussian input distribution, and we can  derive an expression for the disorder+Gaussian average for $\exp\left (-\beta \sum_{a=1}^n H(x_a )\right)$, i.e. the replicated partition function. We can write
\begin{equation*}
\begin{split}
    p(Q\l 0)=\int DxD\Sigma^{(0)}_{ab}\exp\left(-\frac 12\sum_{ab}\Sigma\l 0_{ab}\left(N\l{0}Q\l 0_{ab}-\sum_i x_{ia}x_{ib})\right)\right)\\
    =\int D\Sigma^{(0)}_{ab}\exp\left(-\frac 12 N\l 0 \left(\tr \Sigma \l 0 Q\l 0+\tr \log \left(I-\Sigma \l 0\right)\right)\right)
\end{split}
\end{equation*}
We can use a saddle-point approximation for $\Sigma \l 0$, getting $\Sigma \l 0=I-Q^{(0)-1}$. Plugging this in gives
\begin{equation*}
    p(Q\l 0)=\exp\left(-N\l 0 \kl{Q\l 0}{I}\right),
\end{equation*}
where $\kl{Q\l 0}{I}$ the Kullback-Liebler (KL) divergence between a multivariate Gaussian with covariance $Q\l 0_{ab}$ and one with covariance $\delta_{ab}$. This can be written
\begin{equation*}
   \kl{Q\l 0}{I}=\frac 12 \left(\tr Q\l 0-\tr \log Q \l 0 -n\right) 
\end{equation*}
Combining everything, we have the following expression for the expected value of the replicated partition function:
\begin{align}
\notag \mathbb E \left( Z_\beta^n\right)&=
\int D^{L+1}\Sigma D^{L+1}Q \exp \left(S[\Sigma,Q]\right)\\  
\notag S[\Sigma,Q]&=\frac 12 \beta^2 N\l L \sum_{ab}Q\l L_{ab}+\sum_{\ell=1}^{L}N\l{\ell}\left(\log \Zsq{\Sigma \l {\ell}}{Q \l {\ell-1}}-\frac 12\sum_{ab}\Sigma\l \ell_{ab}Q\l \ell_{ab}\right)\\
\label{eq:fullPerceptronAction} &-\frac 12 N\l 0 \left(\tr \Sigma \l 0 Q\l 0+\tr \log \left(I-\Sigma \l 0\right)\right).
\end{align}

We note in passing that one might also be interested in the extreme value statistics of the squared magnitude of layer $L$ rather than the statistics of a linear combination. Since the squared magnitude of replica $a$ is just $N\l LQ\l L_{aa}$, the action is the same with the first term replaced with $\beta N\l L\sum_aQ\l L_{aa}$.

\subsection{Saddle-Point Equations for The Replicated Action}
\label{subsec:saddlePoint}
The equations \eqref{eq:fullPerceptronAction} provide an exact expression (up to the saddle point approximation for the distribution of $Q^{(0)}$) for calculating the replicated partition function $\mathbb E \left( Z^n\right)$ for multilayer perceptrons in terms of model hyperparameters like $L$ and the $N$s. But actually evaluating this integral over $2Ln^2$ variables is infeasible (especially when $n$ isn't even an integer). Instead, we derive explicit forms of the \textit{saddle point equations} for the action $S$:
\begin{equation*}
    \frac{\partial S}{\partial Q\l \ell}=0,\qquad 
    \frac{\partial S}{\partial \Sigma\l \ell}=0,\qquad \text{for all $0\leq \ell \leq L$.}
\end{equation*}
Evaluating the derivatives of $S$ with respect to the $Q$s and $\Sigma$s will involve taking derivatives of $\log \Zsq{\Sigma}{Q}$. For this it is useful to define
\begin{equation*}
    \evsq{\bullet}{\Sigma}{Q}\equiv\frac{1}{\Zsq{\Sigma}{Q}}\int \bullet\exp\left(-\frac 12 \sum_{ab} z_a \left(Q_{ab}\right)^{-1} z_{b}-\Sigma_{ab}\sigma(z_a)\sigma(z_b)\right)\frac{d^n z}{\sqrt {\det 2\pi Q }}.
\end{equation*}
In words, $\evsq{\bullet}{\Sigma}{Q}$ is the expectation value of $\bullet$ with respect to a probability density over $n$ variables $z_a$ whose joint density is proportional to the exponential in the integrand. A direct computation then gives 
\begin{align*}
 \frac{\partial \Zsq{\Sigma}{Q}}{\partial Q_{ab}}&=\frac 12 Q^{-1}_{a \mu}\evsq{z_\mu z_\nu-Q_{\mu\nu}}{\Sigma}{Q}Q^{-1}_{\nu b},\qquad \frac{\partial \Zsq{\Sigma}{Q}}{\partial \Sigma_{ab}}=\frac 12 \evsq{\sigma(z_a)\sigma(z_b)}{\Sigma}{Q}.    
\end{align*}
After a bit of algebra, our saddle point equations take the following form: 
\begin{subequations}
\label{eq:allEOMs}
    \begin{align}
        Q \l 0 &=\left(I-\Sigma \l 0\right)^{-1}
        \label{eq:q0EOM}\\
        Q \l \ell&=\evsq{\sigma(z_a)\sigma(z_b)}{\Sigma\l \ell}{Q\l {\ell-1}}\quad \textrm{for }1 \leq \ell \leq L       
        \label{eq:qlEOM}\\
        \Sigma \l \ell&=\frac{N\l{\ell-1}}{N\l \ell}\left(\frac 12 Q^{(\ell)-1}_{a \mu}\evsq{z_\mu z_\nu-Q\l \ell _{\mu\nu}}{\Sigma\l{\ell+1}}{Q\l \ell}Q^{(\ell)-1}_{\nu b}\right)\quad \textrm{for }0 \leq \ell \leq L-1 
        \label{eq:SiglEOM}\\
        \Sigma\l L_{a b}&=\beta^2
        \label{eq:SigLEOM}
    \end{align}
    \end{subequations}

\subsection{The $n\to 0$ Limit and the Zipper Method}
\label{subsec:n0}
In this subsection, we will focus on evaluating the right hand sides of the saddle-point equations in the limit $n\to 0$, especially in the replica-symmetric case where we will write
\[
Q^{(\ell)}_{ab}=c^{(\ell)}\delta_{ab}+q^{(\ell)},\qquad \Sigma \l \ell_{ab}=\Lambda \l \ell\delta_{ab}+\Gamma\l \ell.
\]
From equation \ref{eq:q0EOM} we obtain
\begin{equation*}
    c\l 0=(1-\Lambda \l 0)^{-1},\qquad \frac{q\l 0}{c\l 0}=\frac{\Gamma \l 0}{1-\Lambda \l 0}.
\end{equation*}
Moreover, equation \ref{eq:SigLEOM} directly yields
\begin{equation*}
\begin{split}
    \Lambda \l L=0,\qquad  \Gamma \l L=\beta^2
\end{split}
\end{equation*}

The difficult part, of course, is to evaluate the right hand sides of equations \ref{eq:qlEOM} and \ref{eq:SiglEOM}. We discuss this calculation in appendix \ref{app:Zsq}. Once we have our equations of motion, finding satisfying choices for the $c$s, $q$s, $\Lambda$s and $\Gamma$s is surprisingly straightforward. We use what we call the \textit{zipper method}.
\begin{enumerate}
    \item Initialize all $c$s, $q$s, $\Lambda$s and $\Gamma$s. For small and even intermediate $\beta$s, initializing to zero works. For larger $\beta$ it may be wise to initialize using the solution from a smaller $\beta$ and work recursively.
    \item Use equations \ref{eq:q0EOM}, \ref{eq:qlEOM} to solve for $c\l \ell, q\l \ell$ one at a time, starting with $\ell=0$ and continuing to $\ell=L$.
    \item Use equation \ref{eq:SigLEOM}, \ref{eq:SiglEOM} to solve for $\Lambda\l \ell$ one at a time, starting with $\ell=L$ and continuing backwards to $\ell=0$.
    \item Repeat steps 2 and 3 until the answers start to converge. This usually takes no more than four or five repetitions.
\end{enumerate}

This method is guaranteed to work at high temperatures (small $\beta$), and empirically it seems to converge and agree with experiments even at large $\beta$. One word of caution, however. The zipper method cannot spontaneously break replica symmetry. If one starts with replica-symmetric matrices, the matrices will remain RS at every step in the zipper algorithm, even if an RSB solution exists and is more thermodynamically favored. To deal with this, we implement the system with a very slight amount of RSB, and then see if that amount increases or decreases as we iterate the zipper algorithm. 

Before concluding this section and moving on to solving the analyzing the saddle points for various particular architectures and activation function, we briefly remark in the next section on an apparent incompatibility between overlap matrices satisfying the Parisi Ansatz and overlap matrices that appear \textit{after training in LLMs.}

\subsection{Beyond the Parisi Ansatz: Parallelograms Wanted}
Throughout this work, we assume that the overlap matrices obey the \textbf{Parisi Ansatz}: that is that they obey the hierarchical structure discussed in detail in Appendix \ref{subsec:Hierarchical}. This ansatz implies that the likely states of the Boltzmann distribution obey the so-called ultrametric triangle inequality for distance $\forall ABC, d(A,C)\leq \textrm{max}(d(A,B),d(B,C))$. However, states obeying such a structure are incompatible with some of the most important observations of how concepts are embedded in activation space. One clear example is the famous word-arithmetic identity Man-Woman=King-Queen. Identities like this essentially state that the four relevant words form a parallelogram. Unfortunately, parallelograms cannot obey the ultrametric triangle inequality.

This follows from the fact that for any parallelogram $ABCD$, $AC^2+BD^2=2(AB^2+BC^2)$. From this it follows that $\textrm{max}(AC,BD)^2\geq AB^2+BC^2$. But if $AC^2\geq AB^2+BC^2$, then the inequality is violated. Likewise if $BD^2\geq AB^2+BC^2=DC^2+BC^2$.

Although we cannot scan the entire space of $n\times n$ matrices to prove that the dominant ones satisfy the Parisi ansatz and imply ultrametric structure, we have good reason to think they always do. For instance, one can generalize the famous replicon analysis to study the case of layers of overlaps; we find no instability corresponding to parallelograms in activation space.

\section{Deep Linear Networks}
\label{sec:DLN}
In this section, we specialize the computations from section \ref{sec:replica} to the setting of deep linear networks in which the activation function is $\sigma(t)=t$. This is a generalization of the toy problem solved in earlier in subsection \ref{subsec:toymodel}. Specifically, in subsection \ref{subsec:DLNreplica} we will solve equations \ref{eq:allEOMs} in this case. Then, in subsection \ref{subsec:DLNRMT} we confirm these results for the overlaps by deriving them independently using only simple facts from random matrix theory.

\subsection{Replica Trick}
\label{subsec:DLNreplica}
Note that when $\sigma$ is the identity, $H(x)$ is linear in $x$. Hence, the function $\norm{x}^2/2+\beta H(x)$ will only have one local minimum. This guarantees replica symmetry for the saddles points of the replicated action from \eqref{eq:fullPerceptronAction}. Hence, we can write 
\begin{equation}\label{eq:rsa}
Q^{(\ell)}_{ab}=c^{(\ell)}\delta_{ab}+q^{(\ell)}    
\end{equation}
Moreover, when for linear network, our expression \eqref{eq:ZsqDef} for $\Zsq{\Sigma\l \ell}{Q\l {\ell-1}}$ simplifies to
\begin{equation*}
    \Zsq{\Sigma\l \ell}{Q\l {\ell-1}}^{-1}=\sqrt{\det\lr{ I-\Sigma \l \ell Q \l{\ell-1}}}.
\end{equation*}
Substitute this into \eqref{eq:SiglEOM} allows us to exactly solve for the equation of motion of $\Sigma^{(\ell)}$:
\begin{equation}\Sigma \l \ell=\left(Q\l \ell\right)^{-1}-\left(Q\l {\ell-1}\right)^{-1}.
\label{eq:linearSigmaSolve}
\end{equation}
Plugging equation \ref{eq:linearSigmaSolve} back into equation \ref{eq:fullPerceptronAction} gives us the following action in terms of just the $Q$s:
\begin{equation*}
S=\frac 12 \beta^2 N\l{L} \sum_{ab}Q\l{L}_{ab}-\sum_{i=0}^L N\l{i}KL\left(Q\l{i}|Q\l{i-1}\right).
\end{equation*}
where we define $Q\l{-1}$ to be the identity. Using \eqref{eq:rsa} this action becomes 
\begin{equation}
S/n=\frac 12\beta^2 N\l{L}c\l{L}-\frac 12\sum_{i=0}^L N\l i \left(\frac{c \l i}{c \l {i-1}}-\log \frac{c \l i}{c \l {i-1}}+\left(\frac{q\l i}{c\l i}-\frac{q\l {i-1}}{c\l {i-1}}\right)\left(\frac{c \l i}{c \l {i-1}} -1\right)\right).
\label{eq:linearLong}
\end{equation}
where $c \l{-1}\equiv 1$, $q\l{-1}\equiv 0$. Re-computing the new equations of motion for $q\l i$ shows that they are only satisfied if $\frac{c \l i}{c \l {i-1}}=1$ for all $i$, meaning all $c \l i$s are equal to 1 (yes, we just used the equations of motion for $q$ to solve for $c$). We can now differentiate with respect to the $c \l i$s and substitute $c^{(i)}=1$ to get that for $0 \leq i <L$ 
\begin{equation*}
N \l i (q \l i-q \l {i-1})=N \l {i+1} (q \l {i+1}-q \l {i}),
\end{equation*}
while when $i=L$ we have
\begin{equation*}
\beta^2=q \l L-q \l {L-1},
\end{equation*}
which means $q\l i-q\l{i-1}=\frac{N \l L}{N \l i}\beta^2$ for all $i\geq 0$, with $q\l{-1}=0$ by definition. We thus find 
\begin{equation}
c \l i=1,\qquad q \l i=\beta^2\sum_{j=0}^i\frac{N \l L}{N \l j}
\label{eq:DLNRepResults}
\end{equation} 
for all $0\leq i \leq L$. Let's take a moment to translate equation \ref{eq:DLNRepResults} into words. As we go deeper into the network, the (thermodynamically) random part of the activation vectors (parameterized by the $c$s) doesn't change, but the deterministic part (parameterized by $q$) gets longer and longer, making the angle between activation vectors of different replicas ($\arccos \lr{q \l i}/\lr{c\l i+q\l i}$) smaller and smaller.

\subsection{RMT Analysis}
\label{subsec:DLNRMT}
In the special case of Deep Linear Networks, we don't actually need the full power of the replica trick to solve for $Q\l \ell$. Our Hamiltonian is 
\begin{equation*}
H=\left(\prod_{\ell=1}^{L+1} W \l \ell\right)x= \sum_i v_ix_i,
\end{equation*}
for some vector $v$. The distribution of $x$ will be a simple Gaussian with mean $\beta v$ and variance $\textrm{Cov} x_i x_j=\delta_{ij}$. This implies 
\begin{equation*}
    Q\l 0_{ab}=\delta_{ab}+\beta^2\frac 1 {N\l 0} |v|^2.
\end{equation*}
Calculating the expected value of $|v_i|^2$ is straightforward. We express it as
\begin{equation*}
    |v|^2=\sum_{i_1,i_2,...} W\l{L+1}_{i_1}W\l{L}_{i_1i_2}\dots W\l 2_{i_{L-1}i_{L}} W\l 1_{i_Li_{L+1}}W\lT 1_{i_{L+1}i_{L+2}}W\lT 2_{i_{L+2}i_{L+3}}\dots W\lT L_{i_{2L}i_{2L+1}}W\lT {L+1}_{i_{2L+1}i_{2L+2}},
\end{equation*}
where each of the entries is a Gaussian. Performing the Wick Contractions gives us
\begin{equation*}
    |v|^2=\sum_{i_1,i_2,...} \wick{\c4 W\l{L+1}_{i_1}\c3W\l{L}_{i_1i_2}\dots \c2 W\l 2_{i_{L-1}i_{L}} \c1 W\l 1_{i_Li_{L+1}}\c1 W\lT 1_{i_{L+1}i_{L+2}}\c2 W\lT 2_{i_{L+2}i_{L+3}}\dots \c3W\lT L_{i_{2L}i_{2L+1}}\c4 W\lT {L+1}_{i_{2L+1}i_{2L+2}}},
\end{equation*}
The total number of non-vanishing contractions is $\prod_{\ell=0}^L N\l \ell$, corresponding to $i_1=i_{2L+1}, i_{2}=i_{2L+1}... i_{L+1}=i_{L+2}$. Each contraction has a magnitude of $\prod_{\ell=0}^{L-1} (N\l \ell)^{-1}$, for a total value of 
\begin{equation*}
    |v|^2=N\l L,\qquad 
    Q_{ab}=\delta_{ab}+\beta^2\frac{N\l L}{N\l 0}.
\end{equation*}
Calculating $Q\l \ell$ for later $\ell$s requires slightly more complicated Wick patterns. For instance 
\begin{equation*}
    Q\l 1_{ab}=\delta_{ab}+\beta^2 \frac 1 {N\l 1} |W\l 1 v|^2.
\end{equation*}
We can write out
\begin{equation*}
        |W\l 1 v|^2=\sum_{i_1,i_2,...} W\l{L+1}_{i_1}W\l{L}_{i_1i_2}\dots W\l 2_{i_{L-1}i_{L}} W\l 1_{i_Li_{L+1}}\textcolor{red}{W\lT 1_{i_{L+1}i_{L+2}}W\l 1_{i_{L+2}i_{L+3}}}W\lT 1_{i_{L+3}i_{L+4}} W\lT 2_{i_{L+4}i_{L+5}}\dots W\lT L_{i_{2L+2}i_{2L+3}}W\lT {L+1}_{i_{2L+3}i_{2L+4}}
\end{equation*}
There are three ways of handling the Wick contractions among the four innermost $W$s: $\wick{\c1 W\l 1\c1 W\lT 1\c1 W\l 1\c1 W\lT 1}$, $\wick{\c2 W\l 1\c1 W\lT 1\c1 W\l 1\c2 W\lT 1}$ and $\wick{\c1 W\l 1\c2 W\lT 1\c1 W\l 1\c2 W\lT 1}$. All of these contractions have magnitude $\frac{1}{(N\l 0)^2}$, but the combinatorics is different for each contraction. The first one has $(N\l 0)^2$ terms, the second has $N\l 1 N\l 0$ and the third has $N\l 0$. This works out to
\begin{equation}
    Q\l 1_{ab}=\delta_{ab}+\beta^2\left(\frac{N\l L}{N\l 0}+\frac{N\l L}{N\l 1}\right)+O(N^{-1}).
\end{equation}
For $Q\l 2_{ab}=\delta_{ab}+\beta^2\frac{1}{N\l 2} |W\l 2W\l1 1 v|^2$, there will be a total of 9 Wick contractions, of which only 3 will matter at leading order in the $N$s: $\wick{\c4 W\l 2 \c3 W\l 1 \c2 W\lT 1 \c1 W\lT 2 \c1 W\l 2 \c2 W\l 1 \c3 W\lT 1 \c4 W\lT 2}$, 
$\wick{\c2 W\l 2 \c1 W\l 1 \c1 W\lT 1 \c1 W\lT 2 \c1 W\l 2 \c1 W\l 1 \c1 W\lT 1 \c2 W\lT 2}$, 
and $\wick{\c2 W\l 2 \c1 W\l 1 \c1 W\lT 1 \c2 W\lT 2 \c2 W\l 2 \c1 W\l 1 \c1 W\lT 1 \c2 W\lT 2}$.  All have magnitudes $\frac{1}{(N\l 0 N\l 1)^2}$, with multiplicities $N\l 0 (N\l 1)^2 N\l 2$, $(N\l 0)^2 N\l 1 N\l 2$ and $(N\l 0 N\l 1)^2$, giving $q \l 2= \frac{N\l L}{N\l 0}+\frac{N\l L}{N\l 1}+\frac{N\l L}{N\l 2}$. In general, counting up all the non-crossing contractions will let us recover equation \eqref{eq:DLNRepResults}.

\section{Deep Networks With Shaped Activation}
\label{sec:shaped}

\begin{figure}
    \centering
    \includegraphics[width=0.4\linewidth]{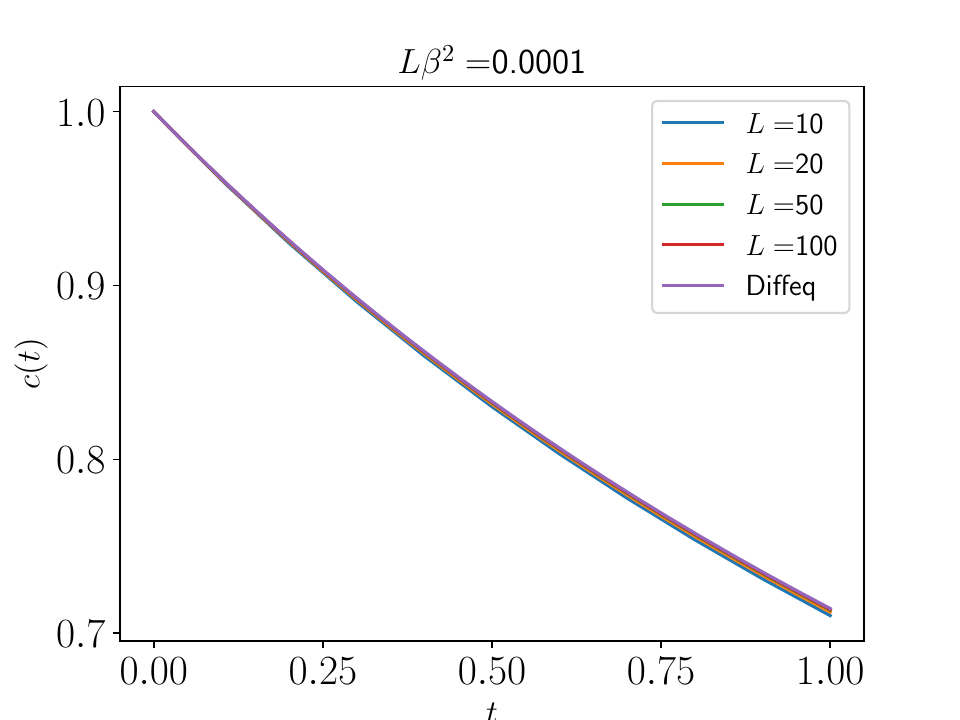}
    \includegraphics[width=0.4\linewidth]{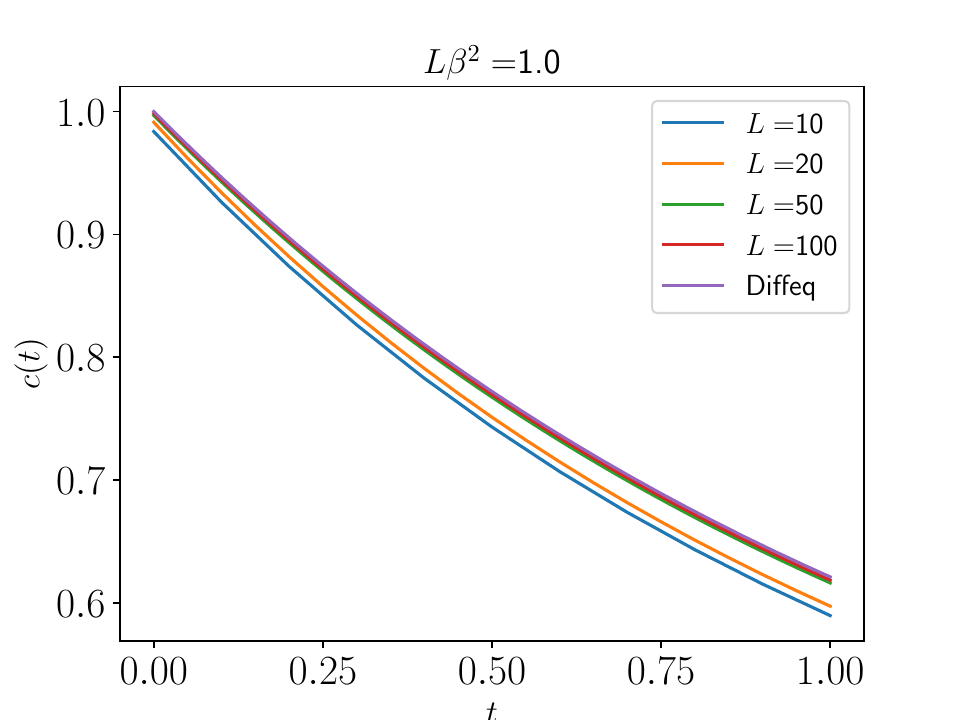}
    \includegraphics[width=0.4\linewidth]{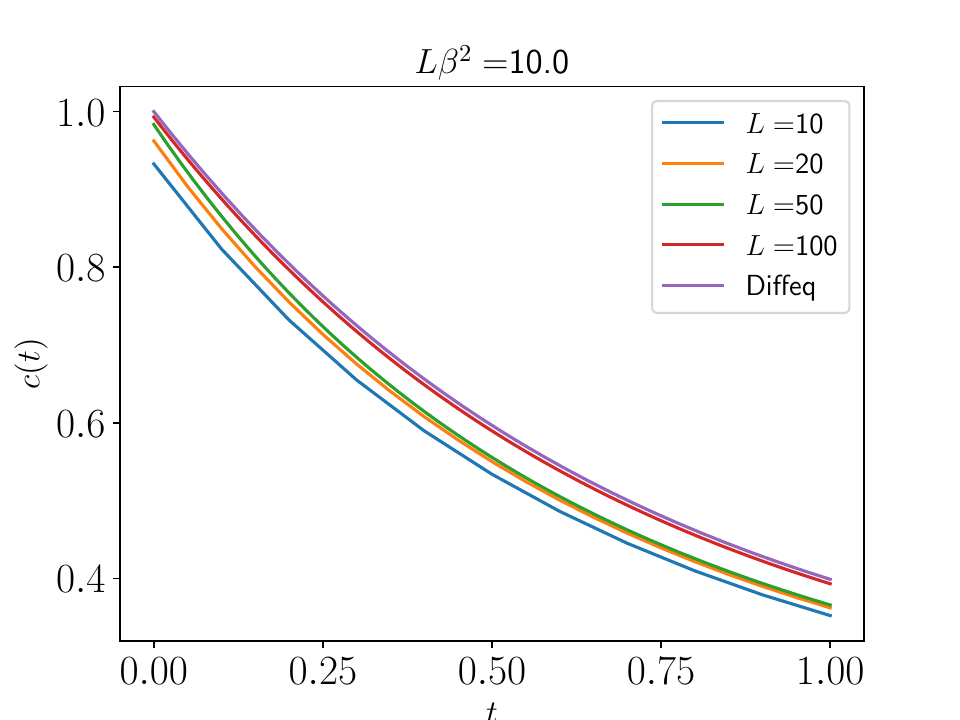}
    \caption{For $\psi=-0.2$, $c(t)$ decreases with depth. The final value of $c$ also decreases as $\beta^2$ increases. Since $c$ can be interpreted as the replica-to-replica variance in the position, we see that as the system gets colder the activations cluster more and more closely around the global minimum.}
    \label{fig:shapedODE}
\end{figure}
In this section we will deal with networks of depth $L$ having a shaped activation
\begin{equation}
    \sigma(z)=z+\frac{\psi z^3}{3L}.
\end{equation}
Such activations have been studied \cite{martens2021rapid,li2023neuralcovariancesdeshaped,noci2023shapedtransformerattentionmodels,zhang2022deeplearningshortcutsshaping,hanin2024bayesianinferencedeepweakly} as a model of deep MLPs which retain reasonable behavior in the large depth limit.

The temperature regime we consider sends the depth $L$ to infinity while holding $\beta^2 L$ constant. This is a sensible limit since it keeps the last-layer inter-replica correlator $Q^{(L)}$ constant for the linear case. One important thing to keep in mind is that although we are taking depth $L$ and width $N$ to infinity, the width is going to infinity first, and is much larger than $L$. In the case of shaped activations, we can still evaluate $\Zsq{\Sigma\l \ell}{Q\l {\ell-1}}$:
\begin{align}
\notag \Zsq{\Sigma_{ab}\l \ell}{Q_{ab}\l {\ell-1}}&= \int  \exp\left(\frac 12 \sum_{ab}\Sigma_{ab}\l\ell\sigma(z_a)\sigma(z_b)\right) Dz\\
 \notag   &=\int  \left(1+\frac{2\psi}{3L}\sum_{ab}\Sigma_{ab}\l\ell z_az_b^3\right)\exp\left(\frac 12 \sum_{ab}\Sigma_{ab}\l\ell z_az_b\right) Dz+O(L^{-2})\\
    &=\sqrt{\frac{1}{\det \lr{I-\Sigma \l \ell Q \l{\ell-1}}}}\left(1+2\frac{\psi}{L}\left(\sum_{ab}\Sigma_{ab}\l\ell \evsq{z_az_b}{\Sigma\l \ell}{Q\l {\ell-1}}\evsq{z_b^2}{\Sigma\l \ell}{Q\l {\ell-1}}\right)\right)+O(L^{-2})
    \label{eq:ZsqShaped}    
\end{align}

We can use the $\Sigma\l\ell$ equations of motion to replace $\evsq{z_az_b}{\Sigma\l \ell}{Q\l {\ell-1}}$ with $Q\l {\ell-1}$. This is not exactly accurate, since it ignores the contribution of $\Sigma$ to the average. However, we are picking a temperature regime where $\Sigma$ is small, so this can be ignored. With this in mind we find 
\begin{equation*}
    \log \Zsq{\Sigma_{ab}\l \ell}{Q_{ab}\l {\ell-1}}=-\frac 12 \tr \log \left(I-\Sigma \l \ell Q \l{\ell-1}\right)+\frac {\psi}L\sum_{ab}\Sigma \l \ell_{ab}Q_{aa}\l{l-1}Q_{ab}\l{l-1}.
\end{equation*}
At this point we can "integrate out" the $\Sigma^{(\ell)}$s, essentially solving the $\Sigma^{(\ell)}$ equations of motion for fixed $Q^{(\ell)}$ and substituting in this value of $\Sigma^{(ell)}$ to get an action exclusively in terms of the $Q^{(\ell)}$s. Equation \ref{eq:fullPerceptronAction} becomes
\begin{equation*}
S=\frac 12 \beta^2 N\l{L} \sum_{ab}Q\l{L}_{ab}-\sum_{i=0}^L N\l{i}KL\left(Q\l{i}|{\tilde Q\l{i-1}}\right),
\end{equation*}
where we define
\begin{equation*}
    \tilde Q_{ab}=\left(1+\frac {2\psi}LQ_{aa}\right)Q_{ab}
\end{equation*}
Since $Q\l\ell$ and $Q\l{\ell-1}$ will be close to each other (differing by an $O(L^{-1})$) amount, we approximate the KL divergence by the distance in the Fisher information metric. Taking the replica symmetric ansatz and forgetting the $\psi$ term for a moment, we have
\begin{equation*}
    \frac{1}{n}KL\left(Q+dQ|Q\right)=\frac 12 \frac{1}{c^2}(1-\frac{2q}{c})dc^2+\frac 1{c^2}dcdq
\end{equation*}
Restoring the $\psi$ term gives us
\[
\frac{1}{n}KL\left(Q+dQ|Q+\frac{2\psi(c+q)}{L}Q\right)=\frac 12 \frac{1}{c^2}(1-\frac{2q}{c})(dc-A_c/L)^2+\frac 1{c^2}(dc-A_c/L)(dq-A_q/L),
\]
where 
\[
A_c=2\psi (c+q)c,\quad A_q=2\psi (c+q)q
\]
represent the expected drifts of $c$ and $q$ as we go from layer to layer, without postselecting on the last layer. If we now define a continuous analog  $t=\ell/L$ of the layer index, then at large $L$ we can turn the sum over $L$ into an integral over $t$:
\begin{equation*}
    \frac{L}{N}S=\frac 12 \beta^2 L c+\int_0^1 \frac 12 \frac{1}{c^2}\left(1-\frac{2q}{c}\right)(c'-A_c)^2+\frac 1{c^2}(c'-A_c)(q'-A_q)dt.
\end{equation*}
with initial conditions $c(0)=1$, $q(0)=0$. If we make the substitutions $u=q/c$, $v=\ln c$, our action becomes
\begin{equation*}
        \frac{L}{N}S=\frac 12 \beta^2 L e^v+\int_0^1 \frac{1}{2}(v'-A_v)^2+(v'-A_v)u' dt,
\end{equation*}
with $A_v=2(1+u)e^v$ represents the expected flow of $A$ (the expected flow of $u$ is 0). The Euler-Lagrange equations for this action give
\begin{equation}
    \begin{split}
        p_u'=-2\psi e^v(v'-A_v)-2\psi e^v u'\\
        p_v'=-A_v(v'-A_v)-A_vu'
    \end{split}
    \label{eq:ODEEOM}
\end{equation}
for $0< t<1$. Here $p_u$ and $p_v$ are the conjugate momenta to $u$ and $v$
\begin{equation}
\begin{split}
    p_u=v'-A_v\\
    p_v=u'+v'-A_v
\end{split}
\end{equation}

This is where our boundary conditions come into play. Our initial conditions are $u=v=0$. Our end condition is $u''(1)=0, v''(1)-A_v'(1)=\beta^2e^{v(1)}.$ These four conditions are enough to uniquely specify a solution. Figure \ref{fig:shapedODE} shows increasingly deep shaped activations approaching the solution to this ODE.

\begin{figure}
    \centering
    \includegraphics[width=0.4\linewidth]{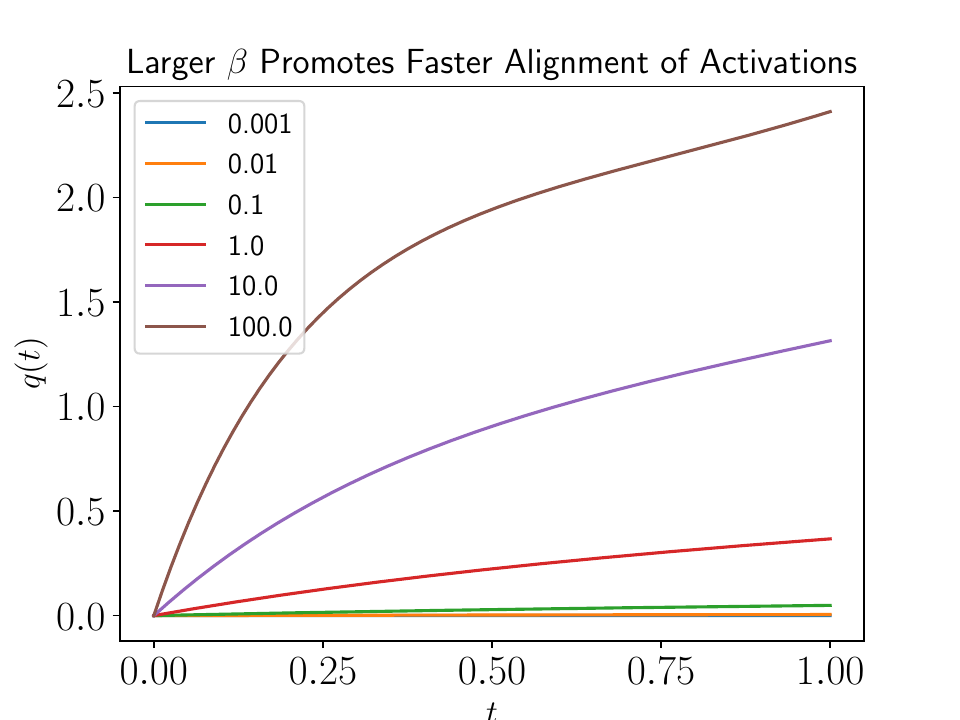}
    \includegraphics[width=0.4\linewidth]{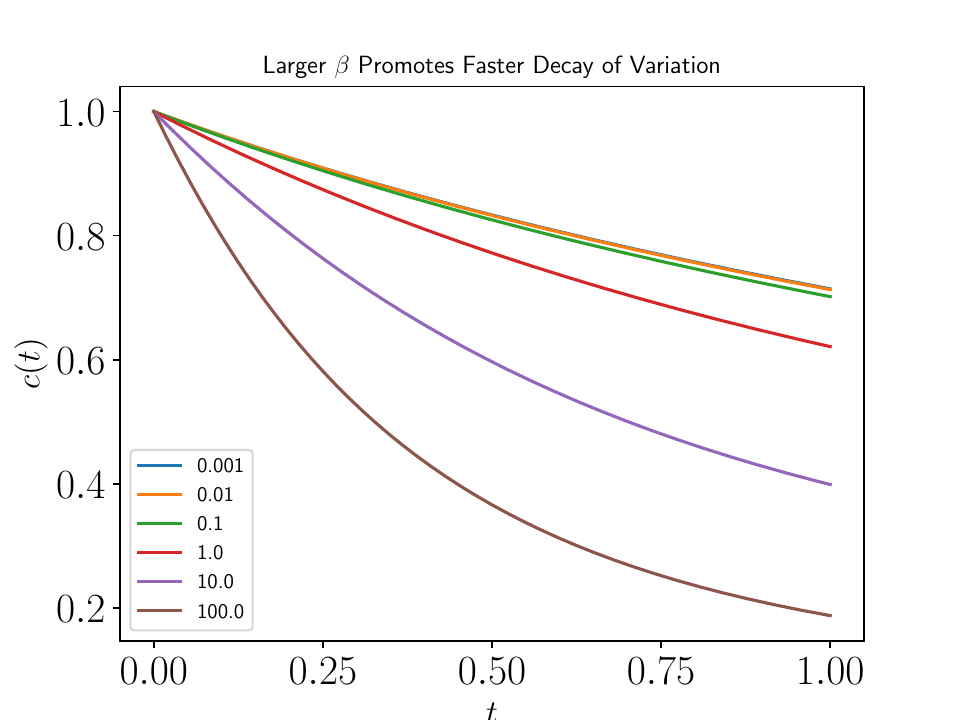}
    \caption{The evolution of overlaps $q(t)$ and variation $c(t)$ in the Gibbs distribution $\exp(-\beta \Htot)$ for deep shaped MLPs.}
    \label{fig:qsTogether}
\end{figure}

At $\beta=0$, the solution of equation \ref{eq:ODEEOM} lines up perfectly with \cite{li2023neuralcovariancesdeshaped, hanin2024bayesianinferencedeepweakly}. Beyond that, however, this work shows how $c$ and $q$ evolve with depth at finite temperature. Figure \ref{fig:qsTogether} shows the solution to \ref{eq:ODEEOM} at various $\beta$. We see that for this negative choice of $\psi$, more positive $\beta$ promotes a faster growth of $q(t)$ and a faster decay of $c(t)$. This growth of $q$ mediated by $\beta$ seems to be universal, a natural consequence of selecting points based on their behavior in the final layer. The $c$ behavior is non-universal and would flip if we flipped the sign of $\psi$.
\section{Thermodynamics and Large-Deviation Principles Multi-Layer Perceptrons}
\label{sec:Numerics}

In this section we present some numerics based on the zipper algorithm described in \S \ref{subsec:n0} and also make several concluding remarks.
\subsection{Evaluating $E(\beta)$}
\label{subsec:ebeta}
Recall that the expected energy at a given temperature is given by
\begin{equation}
    E(\beta)=-\frac{\partial \E{\log Z_\beta}}{\partial \beta},
\end{equation}
Hence, using a saddle point approximation to $\log(Z_\beta)$, we find
\begin{align*}
    n\mathbb E \log {Z_\beta}&=\frac 12 \beta^2 N\l L \sum_{ab}Q\l L_{ab}+\sum_{\ell=1}^{L}N\l{\ell}\left(\log \Zsq{\Sigma \l {\ell}}{Q \l {\ell-1}}-\frac 12\sum_{ab}\Sigma\l \ell_{ab}Q\l \ell_{ab}\right)\\
    &-\frac 12 N\l 0 \left(\tr \Sigma \l 0 Q\l 0+\tr \log \left(I-\Sigma \l 0\right)\right),
\end{align*}
$\Sigma$ and $Q$ are the critical points of the action. Hence, 
\begin{equation*}
    E(\beta)=-\frac 1n\beta N\l L \sum_{ab}Q\l L_{ab}.
\end{equation*}
Given how complicated our expression for free energy is, it is remarkable that energy works out so simply (though, of course, evaluating it means finding the $\beta$-dependent $\Sigma$ and $Q$ which extremize the complicated free energy functional).

\begin{figure}
    \centering
    \includegraphics[scale=0.5]{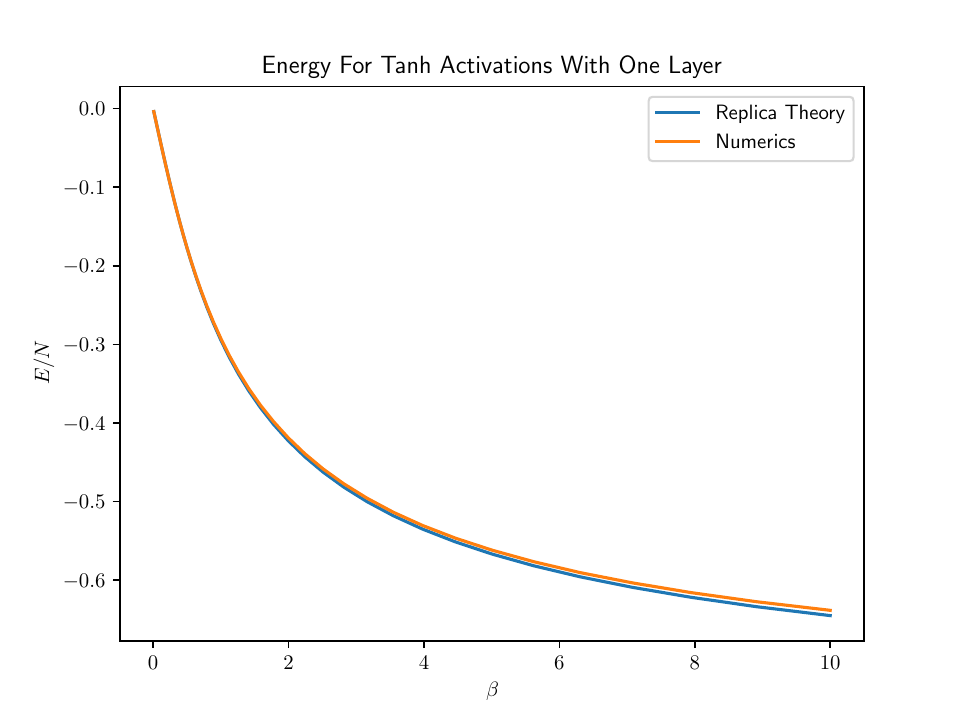}
    \includegraphics[scale=0.5]{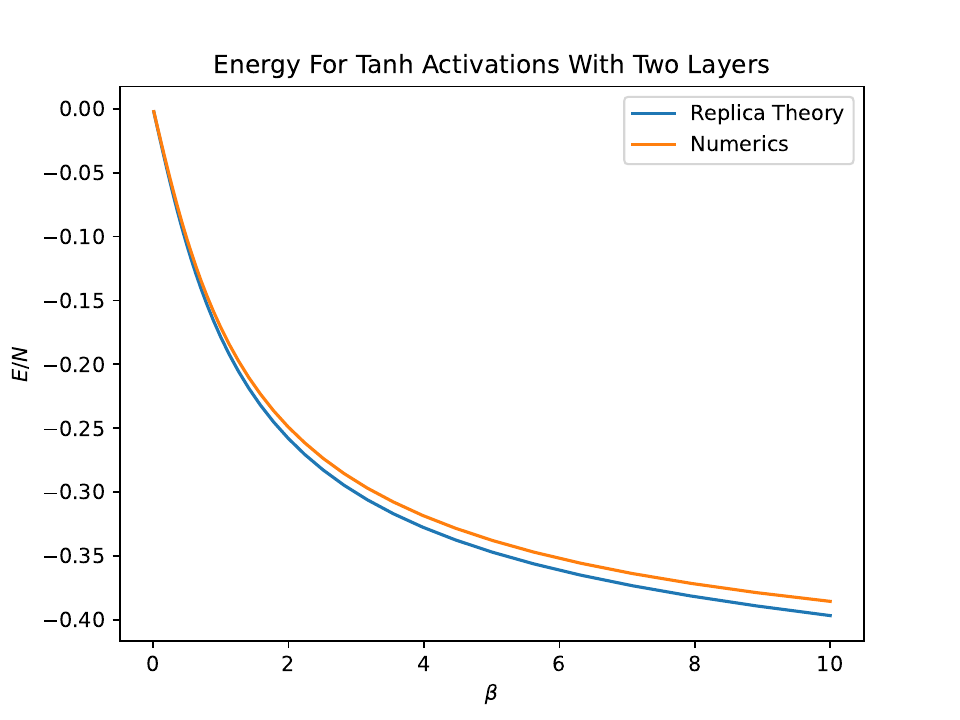}
    \includegraphics[scale=0.5]{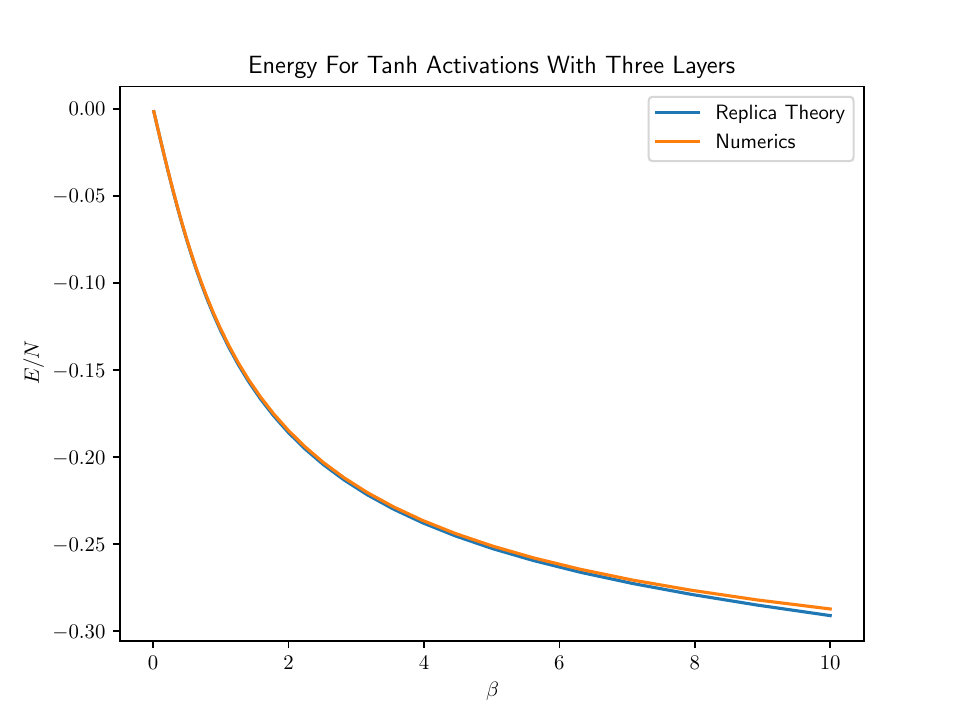}
    \caption{In blue, the replica calculation of the energy at inverse temperature $\beta$. In orange, numerical results obtained by sampling 50 random MLPs of width 200 using stochastic gradient descent.}
    \label{fig:Energies}
\end{figure}

In the case of the deep linear perceptron, $\frac 1n \sum_{ab}Q\l L_{ab}=c\l L=1$, so the energy is precisely $-N\l L \beta.$ We can solve the equations of motion using our zipper algorithm and extract the thermodynamics for a wide variety of hyperparameter choices. Figure \ref{fig:Energies} shows the thermodynamics for models with uniform layer widths, $\tanh$ activations, and a variable number of layers.

Finally, let us comment on how these thermodynamic results can be recast in the language of large deviation theory. In this field, it is common to study the large deviations of some extensive quantity of many variables. In this case the quantity is the output of a multilayer perceptron as a function of the inputs. It is common to find the tails of the probability distribution have the form
\begin{equation}
    P(E)\sim\exp\left(Ns\left(\frac EN\right)\right).
\end{equation}
This fact is just equivalent to the common thermodynamic fact that entropy density and energy density are extensive variables, while entropy and energy scale with $N$.
\subsection{Monte Carlo Sampling}
Deriving a theoretical expression for $E(\beta)$ is all well and good, but we need some numerical calculations to back it up. Outside of a glassy phase, where the low energy space is connected, modern computers can sample from these distributions very efficiently, allowing us to calculate $E(\beta)$. Figure \ref{fig:Energies} compares these numerically extracted energies with the theoretical calculations in subsection \ref{subsec:ebeta}.

\subsection{Replica Symmetry Breaking}
\label{subsec:RSB}

In this subsection, we deal with replica symmetry breaking. Physically, this corresponds to Boltzmann distributions where the measure concentrates into a few valleys of $H$. In terms of the overlap matrices $Q$, it means that instead of being replica symmetric, they take on a hierarchical structure as expounded upon in Appendix \ref{subsec:Hierarchical}.

For fixed $k$ and fixed sequence of $m$s $m_{0\leq i<k}$ we can find a solution to equations \ref{eq:allEOMs} with that pattern of $m$s. Oftentimes, this solution will have $q_0=q_1...=q_k$. Such a solution is replica symmetric even though the ansatz can accommodate RSB.

Figure \ref{fig:RSBGraph} illustrates the presence of replica symmetry breaking for a specific architecture: an activation function of $\sin z$ with one hidden layer. But although it is certainly possible to come up with architectures displaying RSB, many of the most natural choices, including shallow circuits with $\tanh$ or ReLU activations, seem to be replica symmetric.


\section{Conclusion}
\label{sec:Conclusion}

This work is fundamentally about asking a new type of question: what are the statistical behaviors of the special inputs of a neural network. This general question can take many specific forms: here we took `special' to mean `produces of a very negative output' and `a neural network' to mean `an MLP at initialization.' But there are other kinds of special inputs, and other neural network cases of interest. For instance, one could look at the inputs with the largest gradient at the output or the largest magnitude of the final activation layer. More broadly, we believe the techniques we developed here will be of use for studying the statistical mechanics of neural networks in which we treat both weights and data as quenched. 

\subsection*{Acknowledgments}
BH is supported by a 2024 Sloan Fellowship in Mathematics and NSF CAREER grant DMS-2143754 as well as NSF grants DMS-1855684 and DMS-2133806. MW acknowledges DOE grant DE-SC0009988. He would also like to thank Pietro Rotondo and Brian Swingle for useful comments.

\bibliographystyle{ieeetr}
\bibliography{main.bib}
\appendix
\section{Evaluating $\Zsq{\Sigma}{Q}$ in the $n\to 0$ Limit}
\label{app:Zsq}
In this Appendix we tackle the technical problem of evaluating $\Zsq{\Sigma}{Q}$ and $\evsq{f(z_1,z_2)}{\Sigma}{Q}$ for matrices that aren't replica diagonal. Subsection \ref{sub:RSZsq} will deriva an expression for $\Zsq{\Sigma}{Q}$ in the replica symmetric case. Subsection \ref{subsec:Hierarchical} gives a brief overview of Parisi's hierarchical ansatz of replica symmetry breaking matrices, and then explains how to calculate $\Zsq{\Sigma}{Q}$ for such matrices. Subsection \ref{subsec:derivs} calculates the derivatives, of $\Zsq{\Sigma}{Q}$, which are needed to solve the equations of motion, and subsections \ref{subsec:firstOrder}, \ref{subsec:Asymptotic} investigate expansions that work for small $\Sigma$.

\subsection{Replica Symmetric Matrices}
\label{sub:RSZsq}
First, we rewrite
\begin{equation}
\begin{split}
    \Zsq{\Sigma}{Q}=\int \exp\left(-\frac 12 \sum_{ab} z_a \left(Q_{ab}\right)^{-1} z_{b}+\frac 12 \Sigma_{ab}\sigma(z_a)\sigma(z_b)\right)\frac{d^n z}{\sqrt {\det 2\pi Q }}=\\
    \int \exp\left(-\frac 12 \sum_{ab} \left(z_a \left(Q_{ab}\right)^{-1} z_{b}+\zeta_a \left(\Sigma_{ab}\right)^{-1} \zeta_{b}\right)+\sum_{a}\zeta_a \sigma(z_a)\right)\frac{d^n z}{\sqrt {\det 2\pi Q }}\frac{d^n \zeta}{\sqrt {\det 2\pi \Sigma}}
\end{split}
\end{equation}
In other words, the problem of evaluating $\Zsq{\Sigma}Q$ is reduced to evaluating the expected value of $\exp\left(\sum_{a}\zeta_a \sigma(z_a)\right)$ for Gaussian $z$s with covariance $Q$ and Gaussian $\zeta$s with covariance $\Sigma$. We will solve this problem in the case of replica-symmetric $Q,\Sigma$: $Q_{ab}=c\delta_{ab}+q$, $\Sigma_{ab}=\Lambda\delta_{ab}+\Gamma$.

At this point, we define a single Gaussian $\bar z$ with variance $q$ and $n$ iid Gaussian variables $\delta z_a$ with variances $c$. We can just set $z_a=\bar z+\delta z_a$. Analogously, we can define $\zeta_a=\bar \zeta+\delta \zeta_a$, where $\bar \zeta$ has variance $\Gamma$ and the $\delta \zeta$s each have variance $\Lambda$. At this point we can express
\begin{equation}
    \Zsq{\Sigma}{Q}= \int \phi^n(\bar z,\bar \zeta)\frac{\exp \left(-\frac{\bar z^2}{2q}\right)d\bar z}{\sqrt{2\pi q}} \frac{\exp \left(-\frac{\bar \zeta^2}{2\Gamma}\right)d\bar \zeta}{\sqrt{2\pi \Gamma}}
    \label{eq:Znarbitrary}
\end{equation}
with
\begin{equation}
    \phi(\bar z, \bar \zeta)=\int \exp \left((\bar \zeta+\delta \zeta)\sigma(\bar z+\delta z)\right)\frac{\exp \left(-\frac{\delta z^2}{2c}\right)d\delta z}{\sqrt{2\pi c}} \frac{\exp \left(-\frac{\delta \zeta^2}{2\Lambda}\right)d\delta \zeta}{\sqrt{2\pi \Lambda}}
\end{equation}
In the $n\to 0$ equation \ref{eq:Znarbitrary} becomes
\begin{equation}
    \Zsq{\Sigma}{Q}= 1+n\int \log \phi(\bar z,\bar \zeta)\frac{\exp \left(-\frac{\bar z^2}{2q}\right)d\bar z}{\sqrt{2\pi q}} \frac{\exp \left(-\frac{\bar \zeta^2}{2\Gamma}\right)d\bar \zeta}{\sqrt{2\pi \Gamma}}
    \label{eq:Zn0}
\end{equation}
Analogously, $\evsq{f(z_1,z_2)}\Sigma Q$ can be written
\begin{equation}
\begin{split}
    \evsq{f(z_1,z_2)}\Sigma Q=\frac{1}{\Zsq \Sigma Q} \int \mathcal O_f\phi^{n-2}\frac{\exp \left(-\frac{\bar z^2}{2q}\right)d\bar z}{\sqrt{2\pi q}} \frac{\exp \left(-\frac{\bar \zeta^2}{2\Gamma}\right)d\bar \zeta}{\sqrt{2\pi \Gamma}}\\
    \mathcal O_f=\int f(\bar z+\delta z_1,\bar z+\delta z_1)\exp\left((\bar \zeta+\delta\zeta_1)\sigma(\bar z+\delta z_1)+(\bar \zeta+\delta\zeta_2)\sigma(\bar z+\delta z_2)\right)\frac{\exp \left(-\frac{\delta z_1^2+\delta z_2^2}{2c}\right)d^2\delta z}{{2\pi c}} \frac{\exp \left(-\frac{\delta \zeta_1^2+\delta \zeta_2^2}{2\Lambda}\right)d^2\delta \zeta}{{2\pi \Lambda}}
\end{split}
\end{equation}
In the $n\to 0$ limit the first equation simplifies slightly to
\begin{equation}
    \evsq{f(z_1,z_2)}\Sigma Q=\int \mathcal O_f\phi^{-2}\frac{\exp \left(-\frac{\bar z^2}{2q}\right)d\bar z}{\sqrt{2\pi q}} \frac{\exp \left(-\frac{\bar \zeta^2}{2\Gamma}\right)d\bar \zeta}{\sqrt{2\pi \Gamma}}
    \label{eq:fExp}
\end{equation}

Technically, in this section we assumed positive-definiteness for $Q$ and $\Sigma$. This isn't a problem for $Q$, since it is literally the covariance matrix of the activations. But for certain architectures (such as $\sin$ activations) we do indeed find non-PSD solutions for $\Sigma$. The most hygienic way to handle this would be to promote $\zeta$ to a complex variable, and include convolutions in the real direction when we have a positive variance or in the imaginary direction when we have a negative variance. In practice, we dealt with this by replacing our Gausian with a pseudodistribution with negative variance.

\subsection{Hierarchical Matrices}
\label{subsec:Hierarchical}

In this subsection we will give a brief overview of how $\Zsq{\Sigma}{Q}$ can be evaluated for hierarchical matrices obeying the Parisi ansatz. 

To build intuition for the Parisi ansatz, we will first consider the case where the number of replicas $n$ is greater than 1. For a more thorough explanation, we recommend the discussion in chapter 5 of \cite{Parisi_Urbani_Zamponi_2020}. We are also echoing the exposition in \cite{kroeze2024directlyobservingreplicasymmetry}. 
Following the conventions in those sources, we have blocks of size $m_{k}=1$ inside blocks of size $m_{k-1}>1$ inside blocks of size $m_{k-2}>m_{k-1}$ all the way until a block of size $m_{-1}=n$.
The value of $Q_{ab}$ when $a$ and $b$ are in the same block of size $m_{i-1}$ but not the same block of size $m_{i}$ is $q_i$. 
As an example, we will write out the matrix when $m_k=1$, $m_{0}=3$, $m_{-1}=n=6$:
\begin{equation}
    Q_{\alpha\beta}=\begin{pmatrix}
    c+q_1&q_1&q_1&q_0&q_0&q_0\\
    q_1&c+q_1&q_1&q_0&q_0&q_0\\
    q_1&q_1&c+q_1&q_0&q_0&q_0\\
    q_0&q_0&q_0&c+q_1&q_1&q_1\\
    q_0&q_0&q_0&q_1&c+q_1&q_1\\
    q_0&q_0&q_0&q_1&q_1&c+q_1
    \end{pmatrix}.
\end{equation}
The probability of two different replicas being the in the same block and having overlap $q_1$ is $\frac{m_0-1}{n-1}=\frac{3-1}{6-1}=\frac 25$. The probability of being in distinct blocks and having overlaps $q_0$ is $\frac 35$.

If we had a 2-RSB system with $m_2=1$, $m_{1}=3$, $m_{0}=6$, $m_{-1}=n=12$, the overlap distribution would be $p(q_2)=\frac 2 {11}$,$p(q_1)=\frac 3 {11}$,$p(q_0)=\frac 6 {11}$. 

We need to take the $n\to0$ limit to find the free energy, and the rules of the game change when $n<1$. In this case, the ordering of $m$s is flipped: we have blocks of size $m_{k}=1$ inside blocks of size $m_{k-1}<1$ inside blocks of size $m_{k-2}<m_{k-1}$ all the way until a block of size $m_{-1}=n$. This leads to the counterintutive result that at each layer we put larger blocks inside of a smaller block (though only ($\frac{m_{i-1}}{m_{i}}<1$ of then).
Note that overlap probabilities like $\frac{m_0-1}{n-1}$ are still between 0 and 1.

In this subsection, we evaluate the free energy in the RSB scenario. Just as in the replica symmetric case we worked with $\bar z, \bar \zeta$ for the full matrix and $\delta z, \delta \zeta$ for the submatrices, in this case we will define a $z, \zeta$ for each block, sub-block, and sub$^k$-block. We will add in fictitious fields $\tilde z_{ia},\tilde \zeta_{ia}$ where $i$ represents a layer of RSB and the index $\alpha$ runs over all replicas. The $\tilde z$s will be drawn from a joint Gaussian with covariance
\begin{equation}
    \overline{\tilde z_{ia} \tilde z_{jb}}=\delta_{ij} (q_{i}-q_{i-1})*\textrm{Ind}^i_{ab}
\end{equation}
where $\textrm{Ind}^i_{ab}$ is 1 if $a$ and $b$ are in the same block of rank $i$ and zero otherwise (by convention, we are setting $q_{-1}=0, q_{k+1}=c+q_k$). In other words, we are associating an independent $\tilde z$ field with variance $q_{i}-q_{i-1}$ with each block (and analogously for $\tilde \zeta$). Comparing with the RS case, we can identify $\tilde z_{0}$ with $\bar z$, and $\tilde z_{1a}$ with $\delta z_a$. The total value of $z$ for each alpha is given by
\begin{equation}
    z_a=\sum_{i=-1}^k \tilde z_{ia}.
\end{equation}

This can be shown to give total covariance $\overline{z_az_b}=Q_{ab}$
We now define the quantity $Z_i(z_i,\zeta_i)$. This can be interpreted as the partition function of a block of size $m_i$ with effective fields from lower levels totaling $z_i=\sum_{j=0}^{i}\tilde z_j$ and $\zeta_i=\sum_{j=0}^{i}\tilde \zeta_j$. This sort of partition function is a standard quantity in $k$-step RSB analysis \cite{Parisi_Urbani_Zamponi_2020}, but is most often encountered in cases where $Z_i$ depends on only a single fictitious field $X_i$. The generalization to multiple variables was used, for instance, in \cite{kroeze2024directlyobservingreplicasymmetry}. 
We will use $\qb$, $\sigb$ to denote the restrictions of $Q_{ab}-q_i, \Sigma_{ab}-\Gamma_i$ to a block of size $m_i$. In this notation, we can write
\begin{equation}
    Z_i(z_i,\zeta_i)=\int \frac{d^{m_i} zd^{m_i} \zeta}{(2\pi)^{m_i}\sqrt{\det \qb \det \sigb}}\exp \left(\sum_{a,b=1}^{m_i}\zeta_a\sigma(z_a)-\frac 12 \sum_{ab}(z_a-z_i) \left(\qb\right)^{-1} (z_b-z_i)+(\zeta_a-\zeta_i) \left(\sigb\right)^{-1} (\zeta_b-\zeta_i) \right).
\end{equation}

The total $\Zsq{\Sigma}{Q}$ is then $Z_{-1}(0,0)$. We can perform the $2n$-dimensional integral (remembering that $n$ is not an integer) using a recursion. We will have a base case ($Z_k$) and a recursion expressing $Z_{i-1}$ in terms of $Z_i$.

Starting with the $i=k$ case of a single replica per block, we have
\begin{equation}
    Z_k(z_k,\zeta_k)=\int \frac{d^z_a d^\zeta_a}{2\pi\sqrt{c\Lambda}}\exp \left(\zeta_a\sigma(z_a)-\frac 12(z_a-z^k) c^{-1} (z_b-z^ik+(\zeta_a-\zeta_k) \Lambda^{-1} (\zeta_b-\zeta_k) \right)
\end{equation}
Now we derive a recursion for general $Z_{i-1}$. Noting that each block of size $m_{i-1}$ is made out of $\frac{m_{i-1}}{m_i}$ blocks of size $m_i$ and that $z_i,\zeta_i$ are Gaussian random variables with means $z_{i-1}$, $\zeta_{i-1}$ and variance $q_{i}-q_{i-1}$, $\Gamma_{i}-\Gamma_{i-1}$, we have the recursion:
\begin{equation}
    Z_{i-1}(z_{i-1},\zeta_{i-1})=\exp\left[(q_{i}-q_{i-1})\frac{\partial^2}{\partial z_i^2+(\Lambda_{i}-\Lambda_{i-1})}\frac{\partial^2}{\partial \zeta_i^2}\right] \left(Z_{i}(z_i',\zeta_i')\right)^{\frac{m_{i-1}}{m_{i}}}\vline_{z_i',\zeta_i'=z_{i-1},\zeta_{i-1}}.
    \label{eq:diffRecursion}
\end{equation}
In words, to go from blocks of size $m_i$ to $m_{i-1}$ we raise $g_i$ to the power $(m_i-1)/m_i$ and then ``smear'' out the field through a diffusion operator by an amount proportional to $q_i-q_{i-1}$,$\Gamma_i-\Gamma_{i-1}$. 
This gives the recursion relation (equivalent to equation \ref{eq:diffRecursion} except with the Gaussian integrals spelled out in a different way)
\begin{equation}
    Z_{i-1}(z_{i-1},\zeta_{i-1})=\int_{-\infty}^\infty\int_{-\infty}^\infty \frac{dz'_id\zeta'_i}{2\pi\sqrt{(q_{i}-q_{i-1})(\Gamma_{i}-\Gamma_{i-1})}}\exp\left(-\frac{(z'_i-z_{i-1})^2}{2(q_{i}-q_{i-1}})+\frac{(\zeta'_i-\zeta_{i-1})^2}{2(\Gamma_{i}-\Gamma_{i-1}}\right)\left(Z_{i}(z'_i,\zeta'_i)\right)^{\frac{m_{i-1}}{m_{i}}}
\label{eq:intRecursion}
\end{equation}

\subsection{Derivatives of $\Zsq{\Sigma} Q$}
\label{subsec:derivs}
We can use the technology in equation \ref{eq:fExp} to calculate the derivatives of $\Zsq \Sigma Q$ with respect to any individual elements of the $n\times n$ $\Sigma$ or $Q$ matrices. However it is also worth exploring the derivatives with respect to $c,q, \Lambda$ and $\Gamma$ directly. Taking the derivative with respect to $q$ gives us
\begin{equation}
\begin{split}
    \frac{\partial \Zsq{\Sigma}{Q}}{\partial q}= n\int \log \phi(\bar z,\bar \zeta)\left(\frac{\bar z^2-q}{2q^2}\right)\frac{\exp \left(-\frac{\bar z^2}{2q}\right)d\bar z}{\sqrt{2\pi q}} \frac{\exp \left(-\frac{\bar \zeta^2}{2\Gamma}\right)d\bar \zeta}{\sqrt{2\pi \Gamma}}=\\
    \frac n2\int \frac{\partial^2}{\partial \bar z^2}\log \phi(\bar z,\bar \zeta)\frac{\exp \left(-\frac{\bar z^2}{2q}\right)d\bar z}{\sqrt{2\pi q}} \frac{\exp \left(-\frac{\bar \zeta^2}{2\Gamma}\right)d\bar \zeta}{\sqrt{2\pi \Gamma}}
\end{split}
    \label{eq:partialq}
\end{equation}
with $\partial_\Gamma$ analogous. Taking the derivative with respect to $c$ gives us
\begin{equation}
    \frac{\partial \Zsq{\Sigma}{Q}}{\partial c}= n\int \left(\frac{\partial_q\phi(\bar z,\bar \zeta)}{\phi(\bar z,\bar \zeta)}\right)\frac{\exp \left(-\frac{\bar z^2}{2q}\right)d\bar z}{\sqrt{2\pi q}} \frac{\exp \left(-\frac{\bar \zeta^2}{2\Gamma}\right)d\bar \zeta}{\sqrt{2\pi \Gamma}}
    \label{eq:partialc}
\end{equation}
with 
\begin{equation}
\begin{split}
    \partial_q\phi(\bar z,\bar \zeta)=\int \exp \left((\bar \zeta+\delta \zeta)\sigma(\bar z+\delta z)\right)\left(\frac{\delta z^2-c}{2c^2}\right)\frac{\exp \left(-\frac{\delta z^2}{2c}\right)d\delta z}{\sqrt{2\pi c}} \frac{\exp \left(-\frac{\delta \zeta^2}{2\Lambda}\right)d\delta \zeta}{\sqrt{2\pi \Lambda}}=\\
    \frac 12 \int \frac{\partial^2}{\partial \delta z^2}\exp \left((\bar \zeta+\delta \zeta)\sigma(\bar z+\delta z)\right)\frac{\exp \left(-\frac{\delta z^2}{2c}\right)d\delta z}{\sqrt{2\pi c}} \frac{\exp \left(-\frac{\delta \zeta^2}{2\Lambda}\right)d\delta \zeta}{\sqrt{2\pi \Lambda}}
\end{split}
\end{equation}
with $\partial_\Lambda$ analogous.

\subsection{Equation \ref{eq:Zn0} at Small $\Sigma$}
\label{subsec:firstOrder}
When $\beta$ takes on small values, it is useful to understand $\Zsq{\Sigma} Q$ and its derivatives for small $\Sigma$. We will expand to first order in $\Gamma, \Lambda$.

First, we evaluate $\phi(\bar z,\bar \zeta)$ to first order in $\Lambda$ and second order in $\bar \zeta$. We get
\begin{equation}
\begin{split}
    \phi(\bar z,\bar \zeta)\approx \int \left(1+\bar \zeta \sigma(\bar z+\delta z)+\frac 12 \left(\Lambda+\bar \zeta^2 \right) \sigma^2(\bar z+\delta z)\right)\frac{\exp \left(-\frac{\delta z^2}{2c}\right)d\delta z}{\sqrt{2\pi c}}\\
    \log \phi(\bar z,\bar \zeta)\approx \int \bar \zeta \sigma(\bar z+\delta z)\frac{\exp \left(-\frac{\delta z^2}{2c}\right)d\delta z}{\sqrt{2\pi c}}-\frac 12 \left(\int \bar \zeta \sigma(\bar z+\delta z)\frac{\exp \left(-\frac{\delta z^2}{2c}\right)d\delta z}{\sqrt{2\pi c}}\right)^2\\
    +\frac 12\int \left(\frac 12 \left(\Lambda+\bar \zeta^2 \right) \sigma^2(\bar z+\delta z)\right)\frac{\exp \left(-\frac{\delta z^2}{2c}\right)d\delta z}{\sqrt{2\pi c}}
\end{split}
\end{equation}
Yielding
\begin{equation}
    \frac{\Zsq{\Sigma}{Q}-1}n=\frac{1}{2}\left(\Lambda+\Gamma \right)\int \sigma^2(z)\frac{\exp \left(-\frac{z^2}{2(q+c)}\right)d\delta z}{\sqrt{2\pi (q+c)}}-\frac 12 \Gamma \int \left(\int\sigma(\bar z+\delta z)\frac{\exp \left(-\frac{\delta z^2}{2c}\right)d\delta z}{\sqrt{2\pi c}}\right)^2\frac{\exp \left(-\frac{\bar z^2}{2q}\right)d\bar z}{\sqrt{2\pi q}}
\end{equation}
\subsection{The Asymptotic Series in $\Sigma$}
\label{subsec:Asymptotic}
Another strategy for evaluating $\Zsq \Sigma Q$ for small $\Sigma$ is to expand
\begin{equation}
    \exp\left(\frac 12 \sum_{ab}\Sigma_{ab}\sigma(z_a)\sigma(z_b)\right)=1+\frac 12 \sum_{ab}\Sigma_{ab}\sigma(z_a)\sigma(z_b)+\frac 18\sum_{abcd}\Sigma_{ab}\sigma(z_a)\sigma(z_b)\Sigma_{cd}\sigma(z_c)\sigma(z_d)+\dots
\end{equation}
We can termwise evaluate these contributions. For instance the second term has two cases depending on whether $a$ and $b$ are identical or distinct. The first case, which can happen $n$ ways, is an integral over one Gaussian variable $z_a$ with variance $c+q$. The second, which can happen $n(n-1)$ ways is an integral over two variables $z_a,z_b$ with covariance $\begin{pmatrix}c+q&q\\q&c+q\end{pmatrix}$. The sum over $abcd$ needs to be broken into no fewer than 15 different cases, which is already combinatorially difficult and probably more work than the methods at the beginning of this appendix.
\section{Robustness of Results to Training}
One natural question is how robust these results are. For instance, would the slightest amount of training negate them completely? In this appendix, we argue that our results are resilient, and apply to all but an exponentially small fraction of neural nets. We show this by applying a second layer of large-deviation theory.

Recall that our results can be cast in several different languages: they can be read as statements about the thermodynamics of a system with an MLP Hamiltonian, but also as the large deviation theory for a typical MLP taking in Gaussian noise as an input. We showed that for a typical MLP, the volume of space with output $E$ goes as $e^{Ns(E/N)}$, for some definite $s$ we can calculate.
The natural question, then, is to ask about atypical MLPs. In other words, we are calculating the large deviations of our thermodynamic quantities like free energy or entropy. We can probe the deviations for $F(\beta)$ by calculating exponents of $F$. Fortunately, $\mathbb E e^{-n\beta F} =\mathbb E Z(\beta)^n.$ Fortunately, this is the very quantity we already studies with the replica trick, just without the explicit $n\to 0$ limit. We know that for any choice of $n$, the expectation will be some exponential quantity in $N$. This corresponds to a large-deviation principle for $F(\beta)$ with extensive (in $N$) fluctuations being exponentially (in $N$) unlikely. This means that a small amount of training will not be enough to change our results, we need enough training to get to the exponentially small region of MLP-space where our theory doesn't apply.  
\end{document}